\documentclass[12pt]{article}
\pdfoutput=1
\usepackage{sectsty}
\usepackage{mathrsfs}
\allsectionsfont{\sffamily}
\usepackage[nosort]{cite}

\usepackage[pdftex]{graphicx}
\usepackage[font={scriptsize,singlespacing}]{caption}
\usepackage{epstopdf}

\usepackage{amsmath}
\usepackage{amssymb}
\usepackage{subfigure}
\usepackage{hyperref}
\usepackage{url}
\usepackage{xcolor}
\usepackage{color}
\definecolor{amaranth}{rgb}{0.9, 0.17, 0.31}
\definecolor{purple(munsell)}{rgb}{0.62, 0.0, 0.77}
\definecolor{americanrose}{rgb}{1.0, 0.01, 0.24}
\definecolor{palatinateblue}{rgb}{0.15, 0.23, 0.89}
\definecolor{royalblue(web)}{rgb}{0.25, 0.41, 0.88}
\definecolor{hanpurple}{rgb}{0.32, 0.09, 0.98}
\definecolor{beaublue}{rgb}{0.74, 0.83, 0.9}
\definecolor{carminered}{rgb}{1.0, 0.0, 0.22}
\definecolor{brightpink}{rgb}{1.0, 0.0, 0.5}
\definecolor{vividviolet}{rgb}{0.62, 0.0, 1.0}
\hypersetup{ linktoc=all,
    colorlinks, linkcolor={palatinateblue},
    citecolor={brightpink}, urlcolor={amaranth}}

\def\sideremark#1{\ifvmode\leavevmode\fi\vadjust{\vbox to0pt{\vss
 \hbox to 0pt{\hskip\hsize\hskip1em
 \vbox{\hsize2cm\tiny\raggedright\pretolerance10000
 \noindent #1\hfill}\hss}\vbox to8pt{\vfil}\vss}}}%
\newcommand{\bo}{\raise-1mm\hbox{\Large$\Box$}}

\newcommand{\f}[2]{\frac{#1}{#2}}

\newcommand{\y}{\gamma}
\newcommand{\bd}{\boldsymbol}
\newcommand{\la}{\langle}
\newcommand{\ra}{\rangle}

\newcommand{\be}{\begin{equation}}
\newcommand{\ee}{\end{equation}}
\newcommand{\bea}{\begin{eqnarray}}
\newcommand{\eea}{\end{eqnarray}}

\def\brcurs{{\mbox{$\resizebox{.09in}{.08in}{\includegraphics[trim= 1em 0 14em 0,clip]{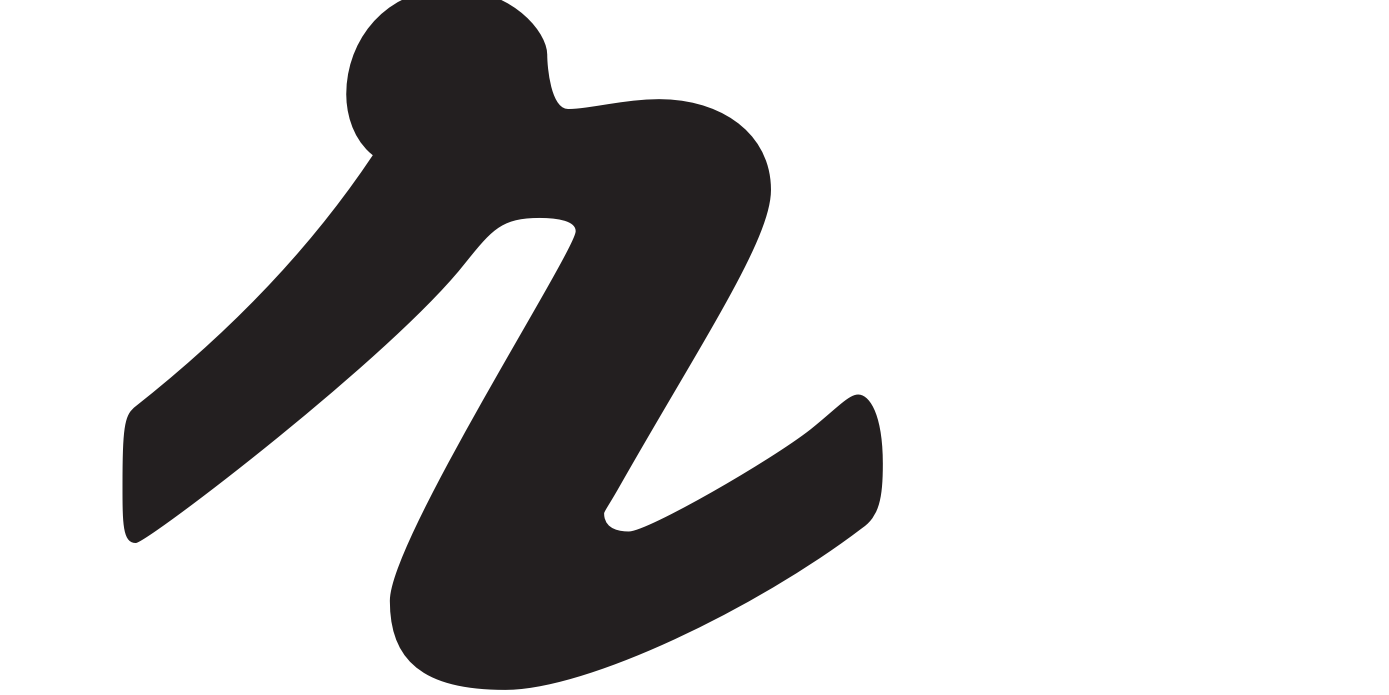}}$}}}

\def\hrcurs{{\mbox{$\hat \brcurs$}}}

\begin{document}
\thispagestyle{empty}
\begin{center}

\null \vskip-1truecm \vskip2truecm

{\LARGE{\bf \textsf{On a Uniformly Accelerated Point Charge moving along a Cusp }}}

\vskip1truecm
\textbf{\textsf{Michael R. R. Good, Thomas Oikonomou, and Gaukhar Akhmetzhanova}}\\
{\footnotesize\textsf{Department of Physics, School of Science and Technology, Nazarbayev University, Astana, Kazakhstan}\\
{\tt Email: michael.good@nu.edu.kz, thomas.oikonomou@nu.edu.kz, gaukhar.akhmetzhanova@nu.edu.kz}}\\





\end{center}
\vskip1truecm \centerline{\textsf{ABSTRACT}} \baselineskip=15pt

\medskip

A uniformly accelerated point charge which moves neither in a straight line nor in a circle, but in a cusp, is investigated.  We find the angular distribution of the Larmor radiation, the constant power, and the intensity in the maximal direction. It is found that the intensity of uniformly accelerated cusp motion scales like non-uniformly accelerated synchrotron radiation. We confirm the exact vacuum excitation spectra of quantized field detectors on the world line.

\vskip0.4truecm
\hrule

\section{Introduction}
Schwinger \cite{Schwinger:1949ym} was first to calculate the angular distribution of power for point charges undergoing linear and circular acceleration.  It was then Letaw \cite{Letaw:1980yv} who found the six classes of uniformly accelerated (stationary) world lines: 
\begin{enumerate}
	\item Inertial $\rightarrow$ Minkowski coordinates \hskip1.0cm 4. Cusp 	 	\item Linear $\rightarrow$ Rindler coordinates \hskip1.75cm 5. Catenary 
 	\item Circular $\rightarrow$ rotating coordinates \hskip1.3cm 6. Helicoid
    
\end{enumerate}
Stationary background systems based on these world lines are of interest because the world lines are trajectories of time-like Killing vector fields.  The lesser known trajectories (4-6) are interesting in their own right and their excitation spectra are connected to the question of coordinate dependence of thermodynamics and quantum field theory in flat spacetime (i.e. they are likely to yield insights into the Unruh effect \cite{Unruh:1976db}). Of these less examined world lines (4-6), the cusp is of particular interest in this article due, in part, to its exactly analytic vacuum spectra. %

There is broad motivation for studying radiation from these stationary trajectories: connections between quantum theory and gravity have been made (e.g. consider relativistic superfluidity \cite{Xiong:2014oga} or geometric creation of quantum vortexes \cite{Good:2014iua}) by investigating the influence of quantum fields under external conditions (like curved spacetime, moving mirrors,\footnote{Consider the accelerated solutions in \cite{Good:2013lca}, and the singularity surviving spin-statistics relation in \cite{Good:2012cp}.} expanding cosmologies, and accelerating world lines). This general line of inquiry continues to lead to investigations that give rise to engaging answers \footnote{Example solutions are explored in the recent black hole-moving mirror correspondence contained in \cite{Good:2016oey}, temperature without a horizon in \cite{Good:2016atu}, and a black hole birth cry and death gasp in \cite{Good:2015nja}.} to challenging questions \footnote{Are black holes springy? \cite{Good:2014uja} Kerr black holes have temperature $2\pi T = g - k$ where $k = m\Omega^2$ is the spring constant and $g= (4m)^{-1}$ is the Schwarzschild surface gravity. }.  

Specifically, accelerated trajectories in flat spacetime and motion in curved spacetime are linked via the equivalence principle.  The `addendum' made to the equivalence principle, (e.g. Unruh effect), is helpful for further understanding the gravitational influence on quantum fields. The cusp motion is an interesting and simple trajectory that warrants more attention, in part, due to the fundamentally important and wide-reaching nature of uniform acceleration.

\subsection{Letaw is almost twice as hot as Planck}
The associated vacuuum fluctuation spectrum for the hyperbolic (straight-line motion) uniformly accelerated world line was found by Unruh to be Planckian, and scales like (using $\omega/\kappa \gg 1$) , 
\be \frac{1}{e^{2\pi \omega/\kappa}-1} \approx e^{-2\pi \omega/\kappa}, \quad \textrm{with} \quad T = \frac{\kappa}{2\pi} = 0.159\kappa.  \ee
The associated vacuum fluctuation spectrum for the cusp uniformly accelerated world line can be said to be Letawian, and Letaw found \cite{Letaw:1980yv} it to have a comparably hotter temperature (by a factor of $\pi/\sqrt{3} = 1.81$), scaling like
\be e^{-\sqrt{12} \omega/\kappa}, \quad \textrm{with} \quad T = \frac{\kappa}{\sqrt{12}} = 0.289\kappa. \ee
 In both cases, the uniform acceleration is $\kappa$, and the spectrum is exactly analytic, which we will confirm.

\subsection{Cusp Motion is Simple }
Besides the Minkowski and Rindler stationary coordinate systems \cite{Rindler:1960zz}, adapted to the inertial and hyperbolic world lines respectively, only the cusp world line has an exactly calculable spectrum in the classification afforded by curvature invariants on which quantized field detectors in a vacuum have time-independent excitation spectra. 

The cusp world line is not only special because its spectrum is exactly calculable, but that it is distinctly non-Planckian, with motion in $2+1$ dimensions with a single parameter, $\kappa$, which describes its proper acceleration and proper angular velocity.  That is, the cusp is a stationary world line solution of the Frenet equations when the curvature invariants of proper acceleration and proper angular velocity, are not only constant but equal to each other. 
Oddly enough, there are a strikingly limited amount of papers that have treated the cusp motion explicitly over the last 35 years: 
\cite{Padmanabhan:1983ub,Takagi:1986kn,Audretsch:1995iw,Rosu:1999ad,Sriramkumar:1999nw,Leinaas:2000mh,Rosu:2005iu,Louko:2006zv,Obadia:2007qf,Russo:2009yd}.

\subsection{Cusp Electromagnetic Lorentz Scalars are Zero}
To further illustrate and motivate the special nature of a charged particle along cusp motion, one may utilize an invariant geometrical description of world lines \cite{Honig} where the Lorentz scalars of homogeneous electromagnetic fields are zero:  

\be \frac{e^2}{m^2}(E^2 - B^2) = \kappa^2 - \tau^2 - \nu^2 = 0, \ee
\be \frac{e^2}{m^2}(\bd{E}\cdot \bd{B}) = -\kappa \nu = 0, \ee
where $\kappa$ (curvature) is the proper acceleration , and $\tau$ (torsion) and $\nu$ (hypertorsion) are components of the proper angular velocity.  For the cusp, $\kappa=\tau$ and $\nu = 0$. It can be further seen that the electric and magnetic fields are perpendicular to each other and their magnitudes are the same.  In the rest of the article, we use units $\hbar = c = k_B = 1$.    

\subsection{Basic Review: Larmor's Formula and Acceleration}\label{sec:acc_basics}

The power from an accelerated point charge is given by relativistic covariant form of Larmor's formula
\be\label{Power1} P = \frac{2}{3} q^2 \alpha^2, \ee
where $\alpha$  is the proper acceleration (see Eq.~(\ref{properacc})) and $q$ is the charge of the electron.  If the proper acceleration is constant, then the power is also  constant.

Consider the difference between the proper acceleration, $\alpha$, and the four-acceleration:  
\be a^{\mu} = \f{d^2 x^{\mu}}{d\tau^2} = \f{dv^{\mu}}{d\tau},\ee
where $v^{\mu} = \f{dx^{\mu}}{d\tau} $ is the four-velocity.  Using the three-acceleration, $\bd{a}=d^2x^i/dt^2$ and the three-velocity $\bd{v} = dx^i/dt$, one may construct the four-acceleration as,
\be a^{\mu} = \f{dv^{\mu}}{d\tau} = \gamma \f{dv^{\mu}}{dt} = \gamma \f{d}{dt} (\gamma, \gamma\bd{v}) = \gamma(\dot{\gamma}, \dot{\gamma}\bd{v} + \gamma \bd{a}) \;,\ee
with $\gamma$ being the Lorentz factor  $\gamma \equiv (1-\beta^2)^{-1/2}$, $\beta\equiv\sqrt{(\bd{v}/c)^2}$ and $\dot{\gamma}\equiv d\gamma/dt$.
The general scalar invariant of this is:
\be a_{\mu}a^{\mu} = \gamma^2\dot{\gamma}^2 - \gamma^2(\dot{\gamma}\bd{v} +\gamma \bd{a})^2 \;,\ee
where we are not necessarily in the instantaneous rest frame of the particle, nor in any other particular reference frame.  Using $\dot{\gamma} = \gamma^3 (\bd{v}\cdot \bd{a})$, as well as $\bd{v}^2 = v^2$, and $\bd{a}^2 = a^2$, the scalar invariant becomes
\begin{subequations}\label{properacc}
\bea \alpha^2 \equiv -a_{\mu}a^{\mu} 
  &=& \y^4a^2 + \dot{\y}^2\\
	&=& \y^4a^2 + \y^6(\bd{v}\cdot \bd{a})^2\\
  &=& \gamma^6a^2 - \y^6(\bd{v} \times \bd{a})^2 \;. \eea
 \end{subequations}
The dot and cross product form are obtained by the use of $\y^2 v^2 + 1 =\y^2$ and $(va)^2 = (\bd{v}\cdot \bd{a})^2 + (\bd{v} \times \bd{a})^2$.  In the case of straight-line motion, parallel vectors $\bd{v}\times \bd{a} = \bd{0}$ yield $\alpha^2 = \y^6 a^2$. In the case of circular motion, the three-acceleration vector is perpendicular to the three-velocity vector, $\bd{v}\cdot \bd{a} = 0$, and one obtains $\alpha^2 = \gamma^4 a^2$.  

Consider the scalar product of the four-acceleration, using the instantaneous rest frame of the particle.  It is straightforward to see how the proper acceleration is constructed from the three-acceleration in the particular frame of rest where $v=0$,
\be a_{\mu}a^{\mu} = (0,\bd{a}) \cdot (0,-\bd{a}) = -\bd{a} \cdot \bd{a} = -a^2 = -\alpha^2 \;,\ee
where $\alpha$ is the acceleration measured in the instantaneous rest frame, i.e., the proper acceleration.

\section{The Cusp}

\begin{figure}[ht]
\begin{center}
\mbox{\subfigure{\includegraphics[width=3.0in]{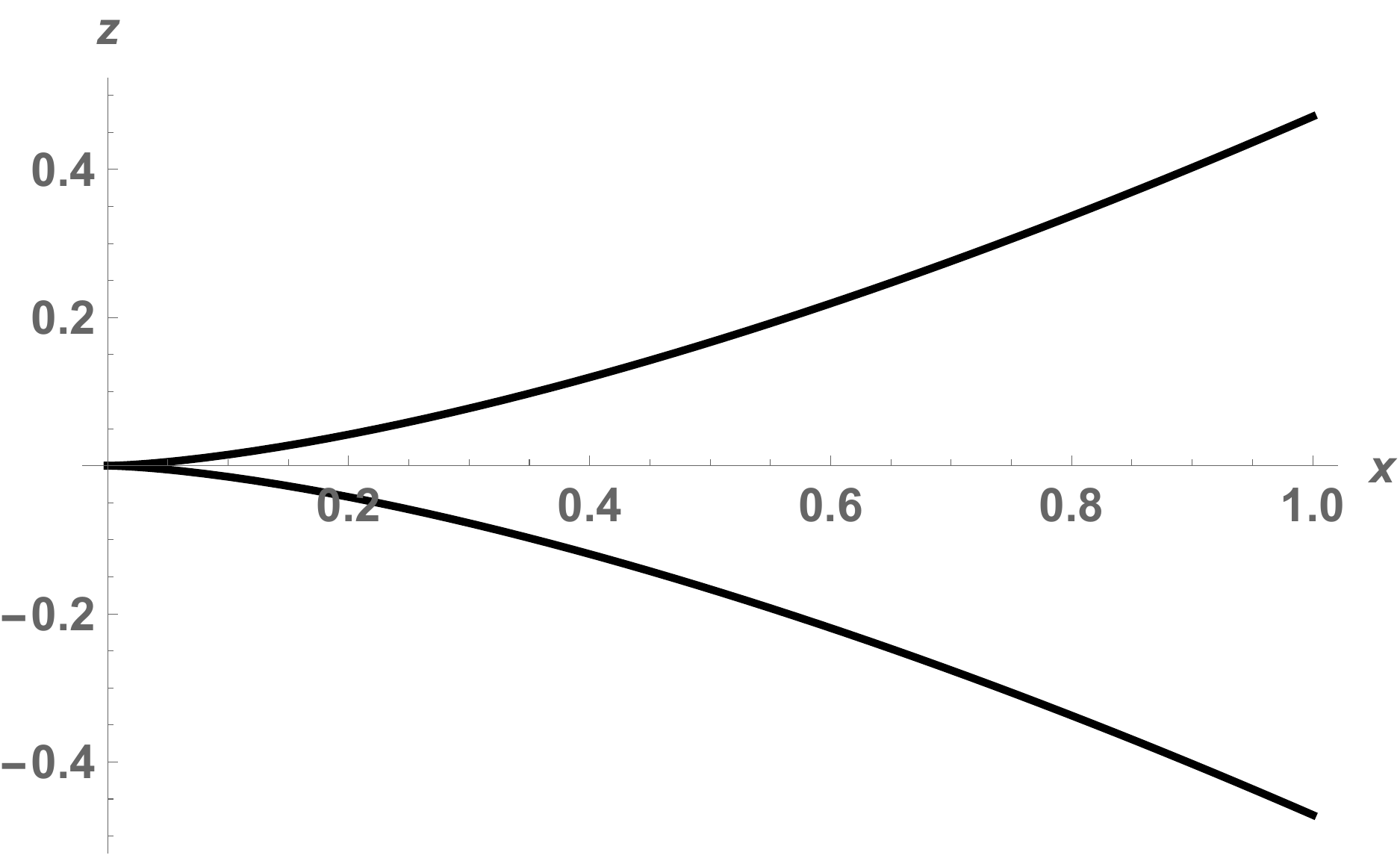}}\quad
\subfigure{\includegraphics[width=3.0in]{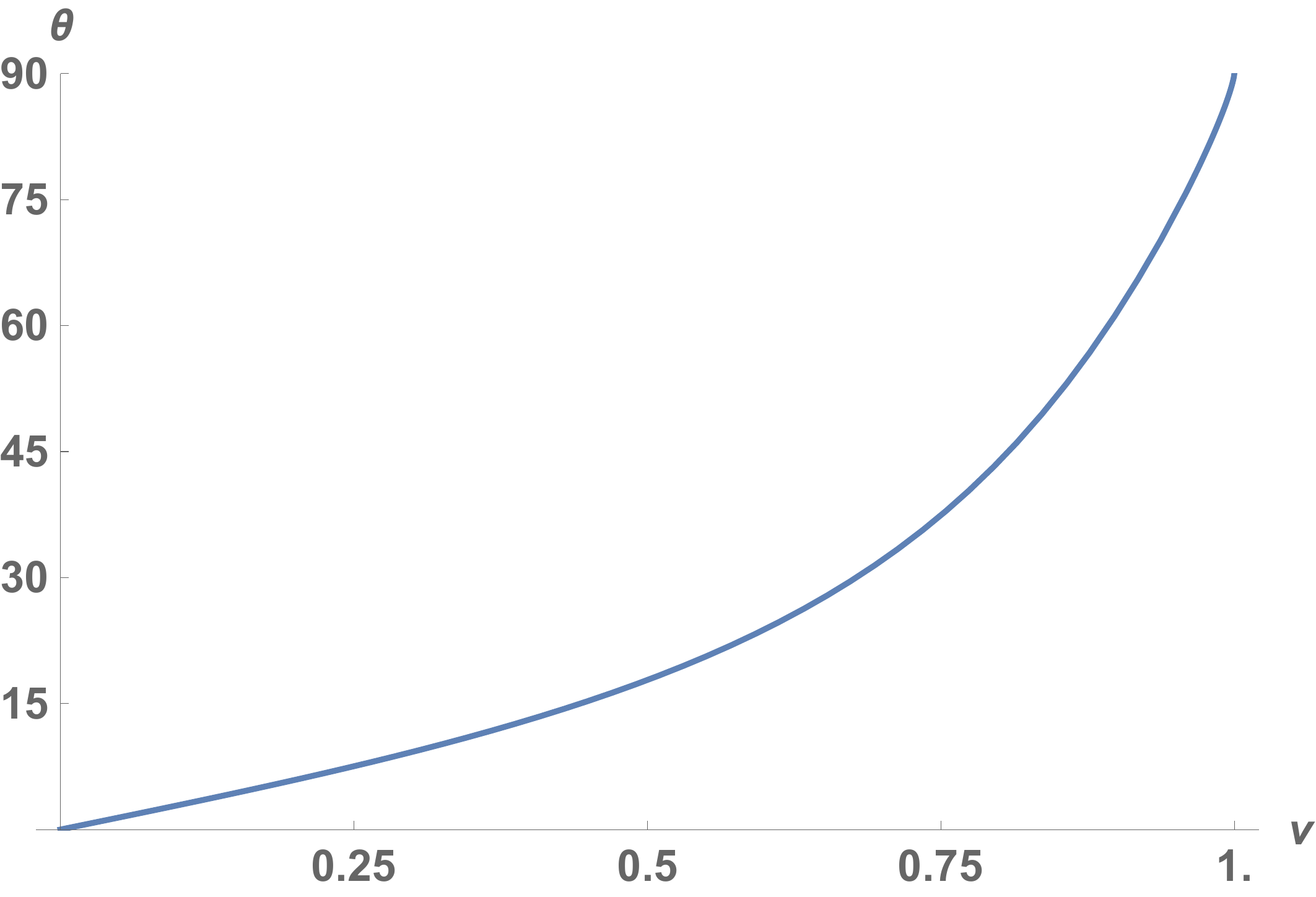} }}
\caption{\label{fig1} \textit{Left}: The cusp projection. \textit{Right}: The angle $\theta$, in degrees, between $\vec{v}$ and $\vec{a}$ of the cusp as a function of speed. } 
\end{center}
\end{figure}

\begin{figure}[ht]
\begin{center}
\mbox{\subfigure{\includegraphics[width=3.0in]{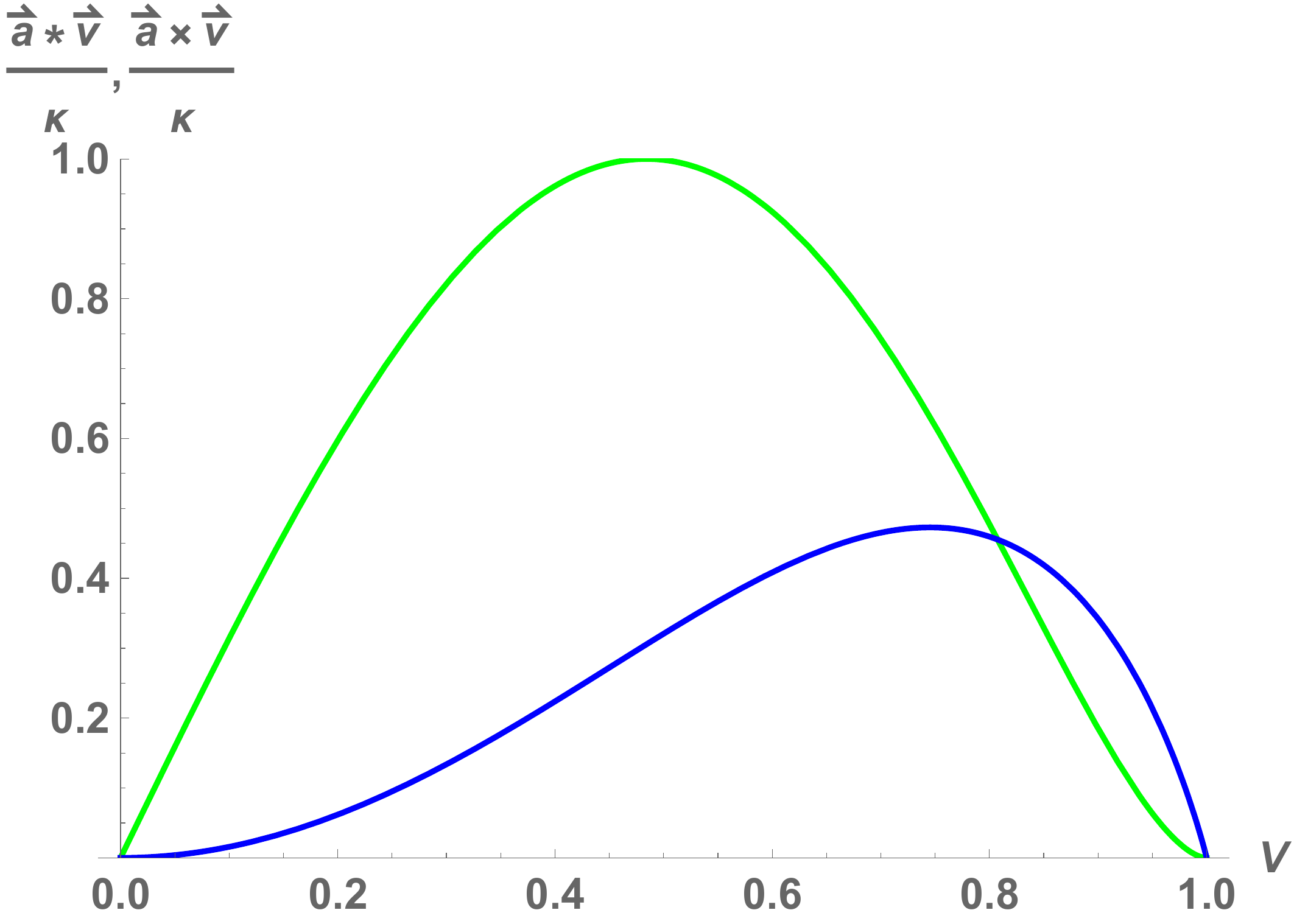}}\quad
\subfigure{\includegraphics[width=3.0in]{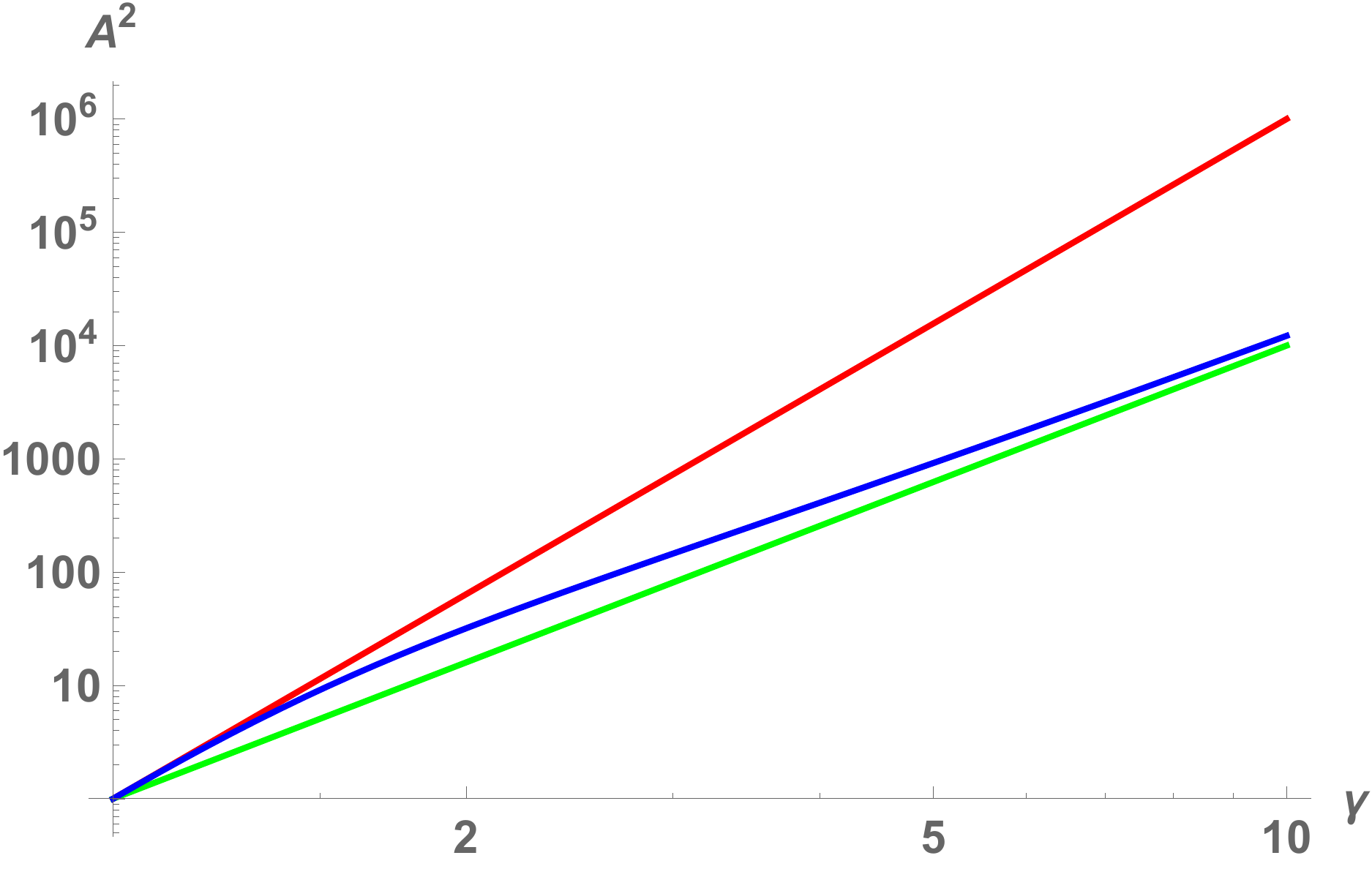} }}
\caption{\label{fig2} \textit{Left}: The dot product and cross product of the the three acceleration with the three velocity.  \textit{Right}: The scaling of the acceleration ratio, $A^2$, with the Lorentz factor, $\gamma$.  Red, blue and green, correspond to the line, cusp, and circular trajectories respectively.} 
\end{center}
\end{figure}

Consider the parametric representation of a timelike world line in flat space,
\be x^\mu (\tau) = \left(\tau + \frac{1}{6} \kappa^2 \tau^3, \frac{1}{2} \kappa \tau^2, 0, \frac{1}{6} \kappa^2 \tau^3\right). \ee
This world line is special class of stationary world lines where the proper angular velocity is equal to the proper acceleration.  To see that the cusp world line is uniform acceleration, consider
\be a^\mu = \frac{d^2}{d\tau^2}x^\mu(\tau) = (\tau \kappa^2, \kappa, 0, \tau \kappa^2), \ee
so that 
\be\label{propacc_constant} 
\alpha^2 \equiv - a^\mu a_\mu = \kappa^2.\ee
One may calculate the proper angular velocity (or torsion) squared by finding the scalar invariant of the four-vector \cite{Rindler:1960zz}
\be \omega^\mu\ = \alpha^{-1} j^\mu - \alpha v^\mu = \left(-\frac{1}{2}\kappa^3 \tau^2, -\kappa^2 \tau, 0, \kappa-\frac{1}{2}\kappa^3 \tau^2\right), \ee 
where $j^\mu = \f{d^3}{d\tau^3}x^\mu$ is the four-jerk and $v^\mu = \f{d}{d\tau}x^\mu$ is the four-velocity. This gives,
\be \omega^2 \equiv - \omega^\mu \omega_\mu = \kappa^2. \ee
The world line has spatial projection in the plane with \footnote{Note the common misprint in Eq. (49) of \cite{Letaw:1980yv} and that of other authors.}
\be z = \frac{\sqrt{2 \kappa}}{3} x^{3/2}, \ee
plotted in Fig.~(\ref{fig1}). In Fig.~(\ref{fig2}) (left), one can view the cross and dot products of the velocity and acceleration.

\subsection{Three Exact Vacuum Excitation Spectrums}
Only the inertial, rectilinear acceleration, and cusp stationary world lines have exactly calculable vacuum excitation spectra, first found by Letaw \cite{Letaw:1980yv} and also emphasized by Rosu \cite{Rosu:1999ad,Rosu:2005iu}.  Using the dimensionless energy, $\epsilon \equiv E/\kappa$, the dimensionless vacuum excitation spectrum on a stationary world line is:
\be S(\epsilon) = \frac{\kappa^2 \epsilon^2}{4\pi^3} \int_{-\infty}^{\infty} \frac{e^{-i \kappa \epsilon\tau}}{ W(\tau,0)} d \tau, \ee
where the geodetic interval is
\be W(\tau,0) = [x_\mu(\tau)-x_\mu(0)][x^{\mu}(\tau)-x^{\mu}(0)], \ee
and the well known exactly analytic results for inertial and linear (Planckian) acceleration are:
\be W_{\textrm{inertial}} = -\tau^{2}, \quad  W_{\textrm{linear}} = \kappa^{-2}(2-2\cosh\kappa \tau), \ee
while for the cusp (Letawian), the result is
\be W_{\textrm{cusp}} = -\left(\tau^2+\frac{\kappa^2 \tau^2}{12}\right)^{-1}. \ee
The inertial spectrum is found from an inverse Fourier transform,
\be S_{\textrm{inertial}}(\epsilon) = -\frac{\kappa^2 \epsilon^2}{4\pi^3} \int_{-\infty}^{\infty} e^{-i \kappa \epsilon\tau}\tau^{-2} d \tau=\frac{\kappa^2 \epsilon^2}{4\pi^3}\pi \kappa \epsilon = \frac{\kappa^3 \epsilon^3}{4\pi^2}. \ee
With a density of states $\kappa^2 \epsilon^2/(2\pi^2)$ the ground state energy per mode is the usual zero point energy: $\kappa \epsilon/2$. The Planckian spectrum is found similarly,
\be S_{\textrm{linear}}(\epsilon) = \frac{\kappa^2 \epsilon^2}{4\pi^3} \int_{-\infty}^{\infty} e^{-i \kappa \epsilon\tau}\frac{\kappa^2}{2-2\cosh\kappa \tau} d \tau = \frac{\kappa ^3 \epsilon ^3}{2 \pi ^2 \left(e^{2 \pi  \epsilon }-1\right)}.\ee 
In the last step the inertial contribution, $\frac{\kappa ^3 \epsilon ^3}{4 \pi ^2}$, is subtracted \cite{Letaw:1980yv}.  The Letawian spectrum is found likewise:
\be S_{\textrm{cusp}} = \frac{\kappa^2 \epsilon^2}{4\pi^3} \int_{-\infty}^{\infty} e^{-i \kappa \epsilon\tau}\left(-\tau^2-\frac{\kappa^2 \tau^2}{12}\right) d \tau = \frac{\kappa ^3 \epsilon ^3}{4 \pi ^2}+\frac{\kappa ^3 \epsilon ^2 e^{-2 \sqrt{3} \epsilon } }{8 \sqrt{3} \pi ^2} = \frac{\kappa ^3 \epsilon ^2 e^{-2 \sqrt{3} \epsilon }}{8 \sqrt{3} \pi ^2}, \ee
where again, the inertial contribution is subtracted in the last step.

\subsection{Acceleration Ratio for the Cusp: Comparison at Low Speeds to Circular and Rectilinear Motions}

For our purpose we rewrite Eq.~(\ref{Power1}) as
\be P_\text{``world line''}=\frac{2}{3}q^2 a^2 A^2_\text{``world line''} \ee
where we denote $A^2_\text{``world line''}$ as the acceleration ratio of a given world line e.g., cusp or linear motion, defined as
\be A^2_\text{``world line''}\equiv\frac{\alpha^2}{a^2}\,, \ee
i.e. the ratio of the squares of the proper-acceleration and the three-acceleration.
Considering circular and linear motion, the former ratio was seen in Sec.~(\ref{sec:acc_basics}) to be equal to
\begin{eqnarray}\label{AccRatio1}
A^2_{\textrm{circ}} = \gamma^4  
\hskip1.0cm\text{and}\hskip1.0cm
A^2_{\textrm{line}} = \gamma^6\,,
\end{eqnarray}
respectively. These simple results beg the question, `what is the acceleration ratio, $A_\textrm{cusp}$?'

To construct the acceleration ratio for the cusp, one examines the non-covariant form by first examining the velocity components, (where $v_y=0$),

\begin{subequations}
\bea v_x &=& \frac{d\tau}{dt} \frac{d x}{d\tau} = \frac{2 \kappa  \tau }{\kappa ^2 \tau ^2+2},\\ 
v_z &=& \frac{d\tau}{dt} \frac{d z}{d\tau} = \frac{\kappa ^2 \tau ^2}{\kappa ^2 \tau ^2+2}.\eea
\end{subequations}
This gives
\be\label{speedSq} v^2 = v_x^2 + v_z^2 = \frac{\kappa ^2 \tau ^2 \left(\kappa ^2 \tau ^2+4\right)}{\left(\kappa ^2 \tau ^2+2\right)^2}\;.\ee
One next examines the components of the acceleration similarly, (where $a_y=0$),
\begin{subequations}
\bea a_x &=& \frac{d\tau}{dt} \frac{d v_x}{d\tau} = \frac{8 \kappa -4 \kappa ^3 \tau ^2}{\left(\kappa ^2 \tau ^2+2\right)^3}\,,\\ 
a_z &=& \frac{d\tau}{dt} \frac{d v_z}{d\tau} = \frac{8 \kappa ^2 \tau }{\left(\kappa ^2 \tau ^2+2\right)^3}.\eea
\end{subequations}

The vector sum is 
\be\label{powerR1} a^2 = a_x^2 + a_z^2 = \frac{16 \kappa ^2 \left(\kappa ^4 \tau ^4+4\right)}{\left(\kappa ^2 \tau ^2+2\right)^6}.\ee
Substituting in Eq. (\ref{powerR1}) the proper time as $\tau=\tau(\gamma)$ through the relation
\be \gamma = 1+\frac{\kappa^2\tau^2}{2}\quad\Leftrightarrow\quad \tau = \kappa^{-1}\sqrt{2\gamma-2} , \ee
obtained  by substituting Eq. (\ref{speedSq}) into the $\gamma$ definition,  we evaluate the acceleration ratio of the cusp motion as
\be \label{cusp_acc_ratio} A^2_{\textrm{cusp}} = \frac{\gamma^6}{\gamma^2 - 2\gamma + 2}. \ee
The behavior of the cusp acceleration ratio for high speeds $\gamma \gg 1$, (low speeds $\gamma \to 1$), is more similar to circular (rectilinear) motion. See Fig.~(\ref{fig2}) (right) for a plot of the $A^2$ values of the cusp, line, and circular ratios. 

It is easy to see in a series expansion of this cusp acceleration ratio, Eq.~(\ref{cusp_acc_ratio}), that at low speeds the Larmor power of cusp motion radiation will more closely scale according to the radiation emitted from straight-line uniformly accelerated motion.  For instance, the acceleration of the point charge moving along a cusp, scales at low speeds,
\be \label{Acusp} A^2_{\textrm{cusp}} \equiv \frac{\gamma^6}{\gamma(\gamma-2)+2} = 1 + 3\beta^2 + 5.75 \beta^4  + \mathcal{O}(\beta)^6, \ee
like linear motion at low speeds,
\be \label{Aline} A^2_{\textrm{line}} \equiv \gamma^6 = 1 + 3\beta^2 + 6 \beta^4  + \mathcal{O}(\beta)^6, \ee
 up to second order, rather than like circular motion at low speeds,
\be \label{Acirc} A^2_{\textrm{circ}} \equiv \gamma^4 = 1 + 2\beta^2 + 3 \beta^4  + \mathcal{O}(\beta)^6.\ee
In light of this low speed scaling, one might assume the angular distribution of the cusp trajectory at low speeds may be more similar to the angular distribution of rectilinear motion, but we shall see this is not the case. 

In this regard, one needs to stress a big difference between the circular and the cusp stationary worldlines; namely, the variation of $\gamma = \gamma(v)$ for the cusp world line.  A circularly uniformly accelerated world line has no change in speed, $\gamma = \textrm{constant}$.  We shall see this come into play when we look at the maximum intensity in the forward direction: i.e. circular motion radiation is always maximum at $\theta^{\textrm{circ}}_{\textrm{max}}=0$ whereas $\theta^{\textrm{cusp}}_{\textrm{max}}=\theta^{\textrm{cusp}}_{\textrm{max}}(v)$, is some function of speed.

\section{Angular Distribution}\label{sec:AD}
Larmor's power formula in Eq. (\ref{Power1}), can be written as,
\begin{eqnarray}\label{Power_2}
P=\int \frac{d P}{d\Omega}\,d\Omega,
\end{eqnarray}
where $dP/d\Omega$ is the angular distribution of radiated power (see Eq. (\ref{AngDistr}) for its definition). Following the methodology in Ref. \cite{Griffiths} the total radiating power in Eq. (\ref{Power_2}) is given by,
\be\label{CuspPower} 
P = \int_0^{2\pi} \int_0^\pi \frac{dP(\theta,\phi)}{d\Omega} \textrm{sin}\theta d\theta d\phi = \frac{2}{3} q^2 \kappa^2, 
\ee
where the angular distribution is,
\be \label{angular1} 
\frac{dP}{d\Omega}=  \frac{2}{3} q^2 \kappa^2 I_{\textrm{cusp}}(\theta,\phi).
\ee
Comparing Eqs. (\ref{Power1}), (\ref{propacc_constant}), and (\ref{angular1}), the function $I_\mathrm{cusp}$ satisfies the relation,
\be \label{unitone} 
\int_0^{2\pi} \int_0^\pi I_{\textrm{cusp}} \; \textrm{sin}\theta d\theta d\phi =1,
\ee
with the explicit expression,
\be\label{even_func} I_{\textrm{cusp}} \equiv 
\frac{\lambda_1
+\lambda_2 \textrm{cos}{\phi} 
+\lambda_3\textrm{cos}^2\phi }{ 
\left( \textrm{cos}\phi + \lambda_4 \right){}^5},\ee
and the $\theta$-factors are:
\begin{subequations}
\begin{eqnarray}
\lambda_0&=&-\frac{3}{8\pi \gamma\sqrt{2\gamma _1}^5 \textrm{sin}^5\theta},\\
\lambda_1&=&\lambda_0
\left[2 + \gamma  \gamma _2 + \gamma _1 \textrm{cos}\theta 
\left(\gamma_3 \textrm{cos}\theta-2 \gamma_2\right)\right],\\
\lambda_2&=& -2\lambda_0 \sqrt{2\gamma _1}\textrm{sin}\theta \left(\gamma _1-\gamma _2 \textrm{cos}\theta\right),\\
\lambda_3&=& 2\lambda_0 \gamma_{3/2}\,\textrm{sin}^2\theta,\\
\lambda_4&=& \frac{\gamma _1 \textrm{cos}\theta - \gamma}{\sqrt{2\gamma _1}\textrm{sin}\theta},
\end{eqnarray}
\end{subequations}
defining $\gamma_n \equiv \gamma-n$ for $n=1,2,3$ and $3/2$.
Eq. (\ref{even_func}) is calculated with straightforward, however effortful vector analysis which is presented in the Appendix.
%

%

\begin{figure}[ht]
\begin{center}
\mbox{\subfigure{\includegraphics[width=3.0in]{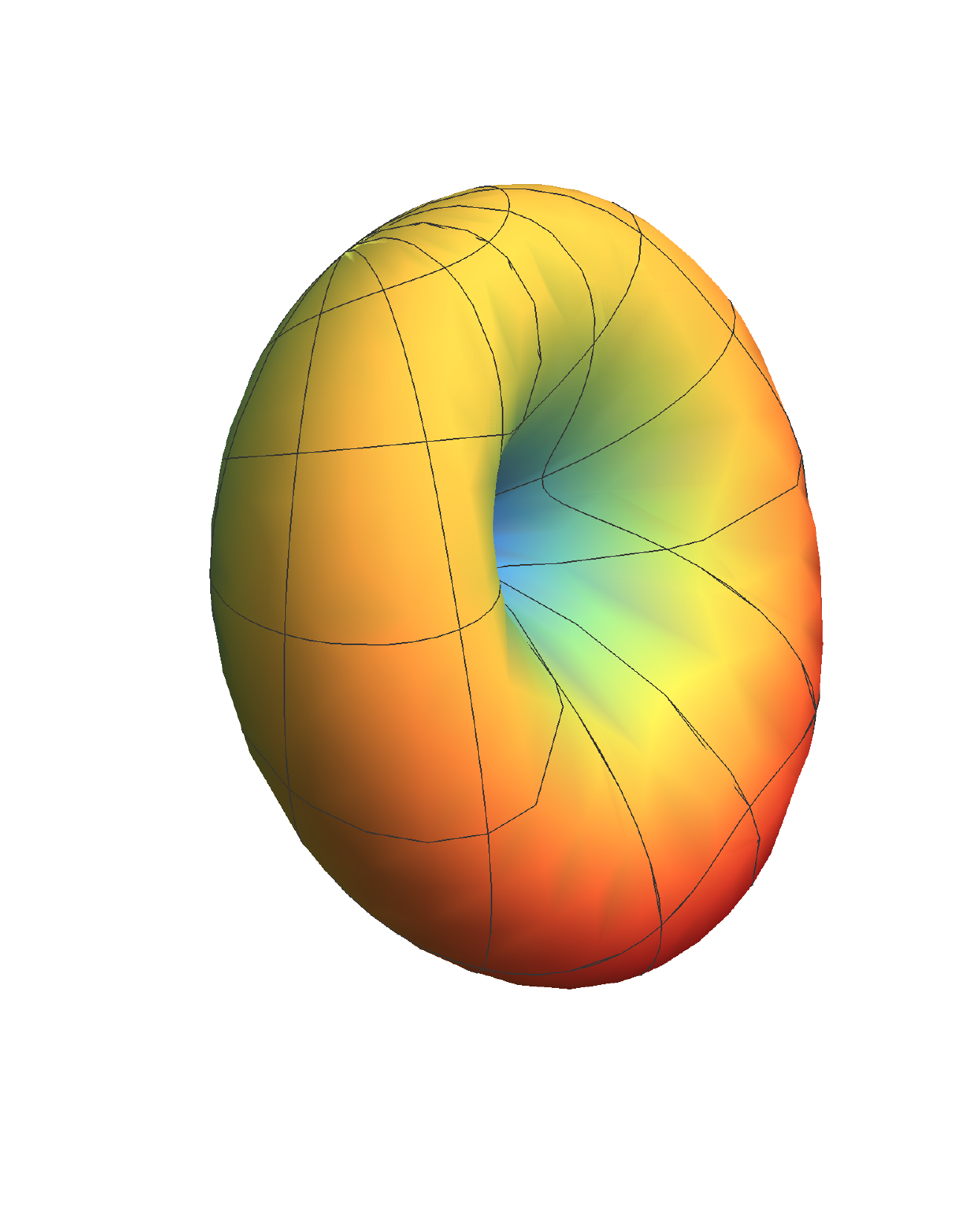}}\quad
\subfigure{\includegraphics[width=3.0in]{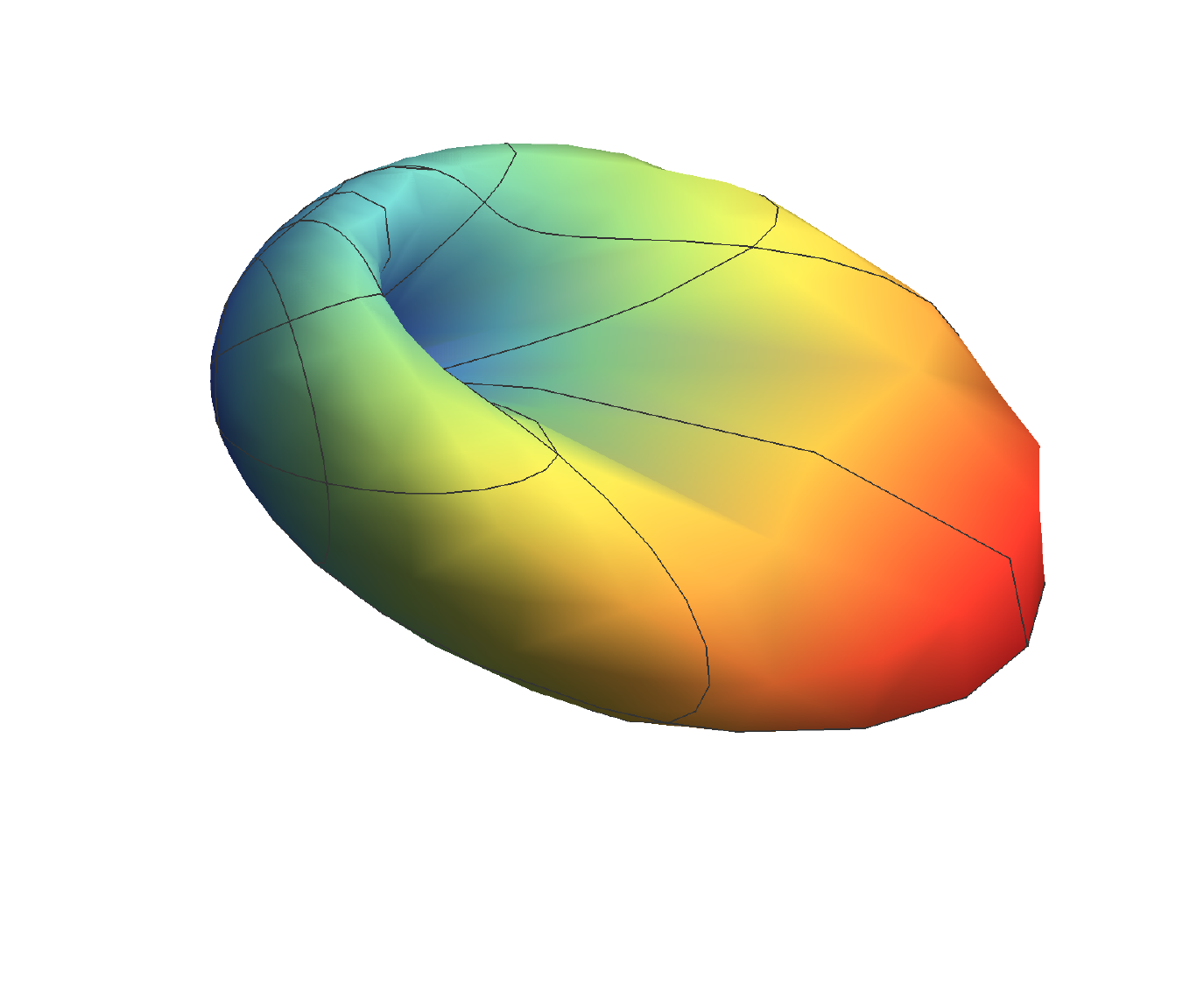} }}
{\caption{\label{fig:333} The angular distribution of cusp radiation at low speeds looks like the usual synchrotron radiation distribution, see \cite{Proceedings} and the 2D plots in the Appendix.  These plots are high speed results of the angular distribution of a uniformly accelerated point charge moving along the cusp trajectory at one-third the speed of light and two-thirds the speed of light, respectively.}} 
\end{center}
\end{figure}

The fulfillment of Eq. (\ref{unitone}) by $I_\mathrm{cusp}$ is confirmed through first an integration of $I_\mathrm{cusp}$ with respect to $\phi$ for $\gamma\geq1$ (or equivalently $\lambda_4\leq-1$) and can be analytically solved.  Defining,
%
\begin{eqnarray}\label{zero_int}
\nonumber
\widetilde{I}_\mathrm{cusp}(\theta)&\equiv&
\int_0^{2\pi}I_\mathrm{cusp}d\phi,
\end{eqnarray}
one may integrate out the $\phi$ dependence,
\begin{eqnarray}
\nonumber
\widetilde{I}_\mathrm{cusp}(\theta)=-\frac{\pi}{4(\lambda_4^2-1){}^{9/2}}
\Big[
\lambda_1\left(3+24\lambda_4^2+8\lambda_4^4\right)-5\lambda_2\lambda_4\left(3+4\lambda_4^2\right)+\lambda_3\left(4+27\lambda_4^2+4\lambda_4^4\right)
\Big].
\end{eqnarray} 
%
Then, integration over $\theta$, using $B \equiv \cos\theta$, yields the unit value as required by Eq.~(\ref{unitone}):
\begin{eqnarray}\label{AngDistr_int}
\int_0^\pi
\widetilde{I}_\mathrm{cusp}(\theta)
\textrm{sin}\theta d\theta 
&=& \int_{-1}^{1}
\widetilde{I}_\mathrm{cusp}(B)
dB=1.
\end{eqnarray}
A high speed and low speed 3D plot of the angular distribution is given in Fig.~(\ref{fig:333}).

%
%

\subsection{Angular Distribution Comparison at High Speeds to Circular and Rectilinear Motions} \label{sec:AD_compare}
We will now compare the angular distributions for straight-line, circular, and cusp uniformly accelerated motions in both the low and high speed regimes in order to get an understanding of the radiation in these limits. For synchrotron angular distribution, $I_{\textrm{circ}}$ is seen to be (see e.g. \cite{Griffiths}),
\be I_{\textrm{circ}} = \frac{3}{8\pi} \frac{(1-\beta\textrm{cos}\theta)^2-(1-\beta^2)\textrm{sin}^2\theta\textrm{cos}^2\phi}{\gamma^4(1-\beta\textrm{cos}\theta)^5}, \ee
and also, like $I_{\textrm{cusp}}$, satisfies an analogous Eq.~(\ref{unitone}).
For low speeds, ($\gamma\to1$, $\beta\to 0$), we saw in \cite{Proceedings} that
\be \label{CircHS} I_{\textrm{circ}} = \frac{3}{8\pi}(1-\textrm{cos}^2\phi\textrm{sin}^2\theta) + \mathcal{O}(\beta), 
\ee
where this was compared to the angular distribution of Larmor radiation of a uniformly accelerated point charge moving along a cusp at low speeds:
\be \label{CuspHS} I_{\textrm{cusp}} = \frac{3}{8\pi}(1-\textrm{cos}^2\phi\textrm{sin}^2\theta) + \mathcal{O}(\beta),\ee
which has exactly the same lowest order term, i.e. for small speeds, cusp and circular motion have the same angular distribution, $I^0_{\textrm{circ}} = I^0_{\textrm{cusp}}$. While for straight-line angular distribution one sees, $ I^0_{\textrm{line}} \neq  I^0_{\textrm{cusp}}$, i.e.
\be \label{LineHS} I_{\textrm{line}} = \frac{3}{8\pi \gamma^6} \frac{\textrm{sin}^2\theta}{(1-\beta\textrm{cos}\theta)^5} =\frac{3}{8\pi} \textrm{sin}^2\theta + \mathcal{O}(\beta). \ee
Interestingly enough, the asymptotic behaviors of $I_\mathrm{circ}$ and $I_\mathrm{cusp}$ are identical for high speeds as well, $\beta \to 1$, to highest order with
\be \label{Icusp} I_{\textrm{cusp}} = -\frac{3}{2\pi}\frac{ (\beta -1)^2}{(\cos\theta -1)^3} + \mathcal{O}(\beta-1)^{9/4},
\ee
and
\be \label{Icirc} I_{\textrm{circ}} = -\frac{3}{2\pi}\frac{ (\beta -1)^2}{(\cos\theta -1)^3} + \mathcal{O}(\beta-1)^{3}.
\ee
The straight-line $I_\mathrm{line}$ is clearly distinguished from the former two angular distributions yielding,
\be \label{Iline} I_{\textrm{line}} = -\frac{3}{\pi}\frac{\sin^2\theta (\beta -1)^3}{(\cos\theta -1)^5} + \mathcal{O}(\beta-1)^{4}.
\ee
%
Comparing the asymptotic behaviors of the angular distributions in Eqs. (\ref{CircHS})-(\ref{LineHS}) with the power expressions in Eqs. (\ref{AccRatio1}) and (\ref{cusp_acc_ratio}) for low speeds we observe that in both measures the results consistently demonstrate a similarity between the circular and the cusp motion. However, what is surprising is that for high speeds these two measures, $P_{*}$ and $I_{*}$ (where $P_{*}\sim\int I_{*} d\Omega$) yield different results. Namely, although $I_\mathrm{cusp}$ and $I_\mathrm{circ}$ asymptotically coincide, as one can see in Eqs. (\ref{Icusp}) and (\ref{Icirc}), their integrals on the other hand do not. In the low speed limit the power $P_\mathrm{cusp}$ behaves similarly to  $P_\mathrm{line}$ rather than $P_\mathrm{circ}$ (see Eqs. (\ref{Acusp})-(\ref{Acirc})).

It is worth remarking at this point that the similarities in the asymptotic expressions between the circular and the cusp calculations above are only symbolic and do not imply that the two aforementioned motions coincide. The reason is that the parameter $\gamma$ (or $\beta$) of the cusp motion varies as a function of speed along the same world-line, while changing $\gamma$ in the circular motion corresponds to a different circular trajectory of different radius.

\section{Intensity of Radiation in the Maximal Direction}
The intensity of maximum radiation in the ultra-relativistic case, ($\gamma \to \infty$), 
\be f \equiv \frac{(\left. d P/d\Omega \right|_{\theta=\theta_{\textrm{max}}})_{\textrm{ultra-rel}}}{(\left.d P/d\Omega\right|_{\theta=\theta_{\textrm{max}}})_{\textrm{rest}}}, \ee
was calculated for rectilinear motion in \cite{Griffiths}.  The angle $\theta_{\textrm{max}}$ at which maximum radiation is emitted occurs when
\be \frac{d}{d\theta} \Big[\frac{dP}{d\Omega}\Big] = 0. \ee
Calculating the derivative of Eq. (\ref{angular1}) with respect to the angle $\theta$ in the plane $\phi = 0$, is not succinct, however it is straightforward. However, determining the correct $\theta_{\max}$ corresponding to the global maximum of the angular radiation, is not trivial due to the plethora of local extrema and very complicated expressions. After numerical and graphical analysis we found the solution which is lengthy.  It is expanded in the high speed limit as, 
\be \label{maxangle}\theta^{\textrm{cusp}}_{\textrm{max}} = 
\sqrt{\frac{2}{\gamma}} + \frac{5}{21\sqrt{2}} \left(\frac{1}{\gamma}\right)^{3/2} + \mathcal{O}\left(\frac{1}{\gamma}\right)^{5/2}. \ee
Using this in the angular distribution and accounting for ultra-relativistic speeds, we find the intensity of maximum radiation from a charged point particle moving along a cusp as
\be \label{maxint} f_{\textrm{cusp}} = 8 \gamma^6. \ee
For the sake of completeness, we present the analogous results for the straight line and the circular motion as well. Precisely, the maximum angle for the aforementioned trajectories are determined to be $\theta_{\textrm{max}}^{\textrm{line}} = (2\gamma)^{-1}$ and $\theta_{\textrm{max}}^{\textrm{circ}} = 0$, respectively, with the corresponding ultra-relativistic intensity,

%
\be 
f_{\textrm{line}} = \frac{8192}{3125} \gamma^8 = 2.62 \gamma^8,
\qquad
f_{\textrm{circ}} = 8 \gamma^6\,.
\ee
%
%
As can be seen here, the intensity measure for cusp and circular world-lines exhibits the same dependence not only in the power of $\gamma$ but also in the proportionality factor in the high speed limit, clearly distinguished from the straight line behavior. It should be stressed however, that the $\gamma$-dependence of $f_{\mathrm{circ}}$ is meaningful only for non-uniformly accelerated motion, since $f_\mathrm{cusp}$ is a function of speed while $f_\mathrm{circ}$ is simply a constant for uniform acceleration. For the intensity of circular motion to scale like $8\gamma^6$, one must abandon the stationary world-line by discarding uniform proper acceleration, resulting in a change of speed $v$.
%

\subsection{Asymmetry Measure} 

The preferred direction of emission at low speeds can also be considered by a simple measure of asymmetry due to Schwinger\cite{Schwinger:1949ym}, who considered the mean value of $\sin^2\theta$,

\be \label{asymmetry1}\la \sin^2\theta \ra = \int \sin^2\theta   \frac{(\left. d P/d\Omega \right)}{P} d\Omega. \ee
It is easy to calculate that
\be \la \sin^2\theta \ra_{\textrm{line}} = -\frac{\left(\beta ^2-1\right) \left(\beta  \left(5 \beta ^2-3\right)+3 \left(\beta ^2-1\right)^2 \tanh ^{-1}\beta\right)}{2 \beta ^5}\,,
\ee
\be \la \sin^2\theta \ra_{\textrm{circ}} = -\frac{\left(\beta ^2-1\right) \left(\beta  \left(\beta ^2+3\right)+3 \left(\beta ^4-1\right) \tanh ^{-1}\beta \right)}{4 \beta ^5}\,,
\ee
and for the cusp motion
\begin{eqnarray}\label{cusp_asmm}
\nonumber
\la \sin^2\theta\ra_{\textrm{cusp}} &=& 
\int_0^\pi
\left[
\int_0^{2\pi}  I_{\textrm{cusp}}(\phi,\theta)d\phi
\right] 
\sin^3\theta d\theta =
\int_0^{\pi}
\widetilde{I}_\mathrm{cusp}(\theta)
\sin^3\theta d\theta \\
\nonumber&=&
\int_{-1}^{1}
\widetilde{I}_\mathrm{cusp}(B)
(1-B^2) dB\\
\nonumber&=&
\frac{1}{8\beta^5\left(1 + \sqrt{1 - \beta^2}\right)^2}
\Big\{
117 \beta - 339 \beta^3 + 351 \beta^5 - 129 \beta^7\\
\nonumber&+&
\beta \sqrt{1 - \beta^2} 
\left(-93 + 212 \beta^2 - 135 \beta^4 + 32 \beta^6\right)\\
&-&
3 \left(1 - \beta^2\right)^2 
\left[39 - 48 \beta^2 + 13 \beta^4 + 
   \sqrt{1 - \beta^2} \left(-31 + 19 \beta^2\right)\right] \tanh^{-1}\beta
\Big\}.
\hskip0.5cm
\end{eqnarray}
%
%
As can be seen, the expression of $\la \sin^2\theta \ra_{\textrm{cusp}}$ is more complicated than $\la \sin^2\theta \ra_{\textrm{circ}}$ and $\la \sin^2\theta \ra_{\textrm{line}}$. At low speeds, for all above mentioned motions, the asymmetry measure yields
\begin{eqnarray}
\la \sin^2\theta \ra_{\textrm{line}}=\frac{4}{5} + \mathcal{O}(\beta)\,,\qquad
\la \sin^2\theta \ra_{\textrm{circ}}=\frac{3}{5} + \mathcal{O}(\beta)\,,\qquad
\la \sin^2\theta \ra_{\textrm{cusp}}=\frac{3}{5} + \mathcal{O}(\beta),
\end{eqnarray}
demonstrating again that at low speeds the asymmetry measure for cusp motion is more closely aligned with circular motion, which is to be expected from the angular distribution results.
%
%
So it should be explicitly clear that even though the total constant cusp power more closely scales to rectilinear power, the cusp angular distribution demonstrates synchrotron character at low speeds.

\begin{figure}[ht]
\begin{center}
\mbox{
\subfigure{\includegraphics[width=1.4in]{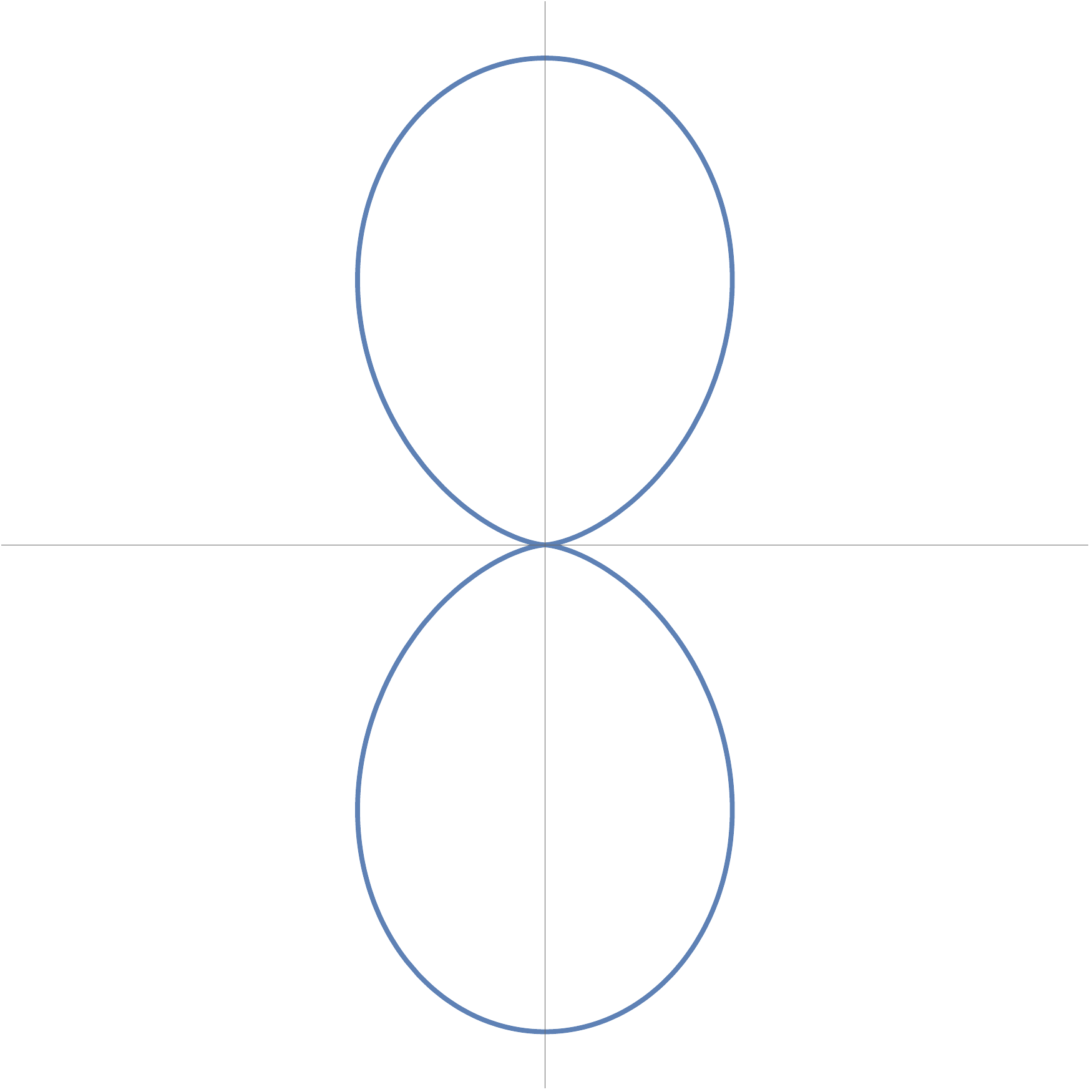}}\quad
\subfigure{\includegraphics[width=1.4in]{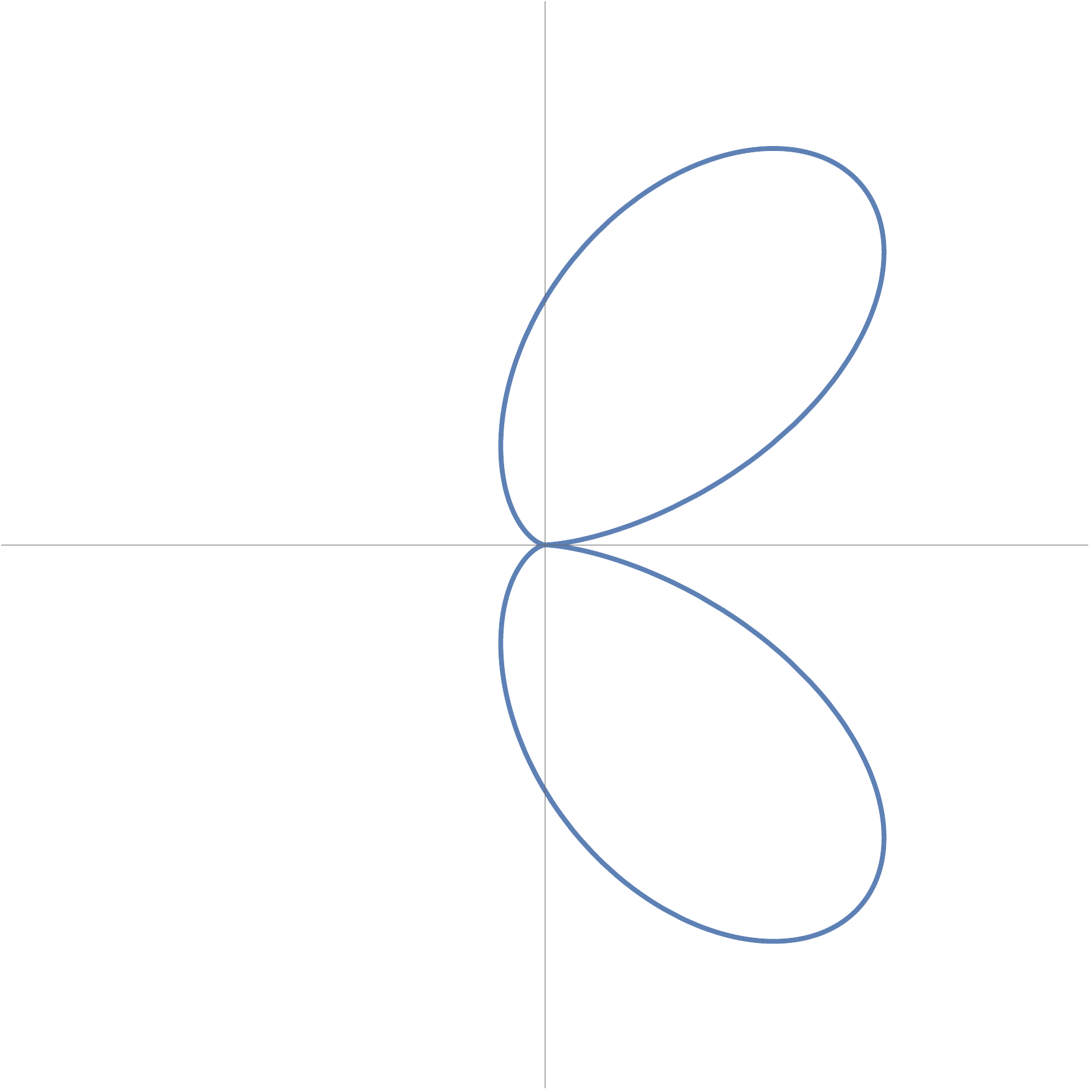}}\quad
\subfigure{\includegraphics[width=1.4in]{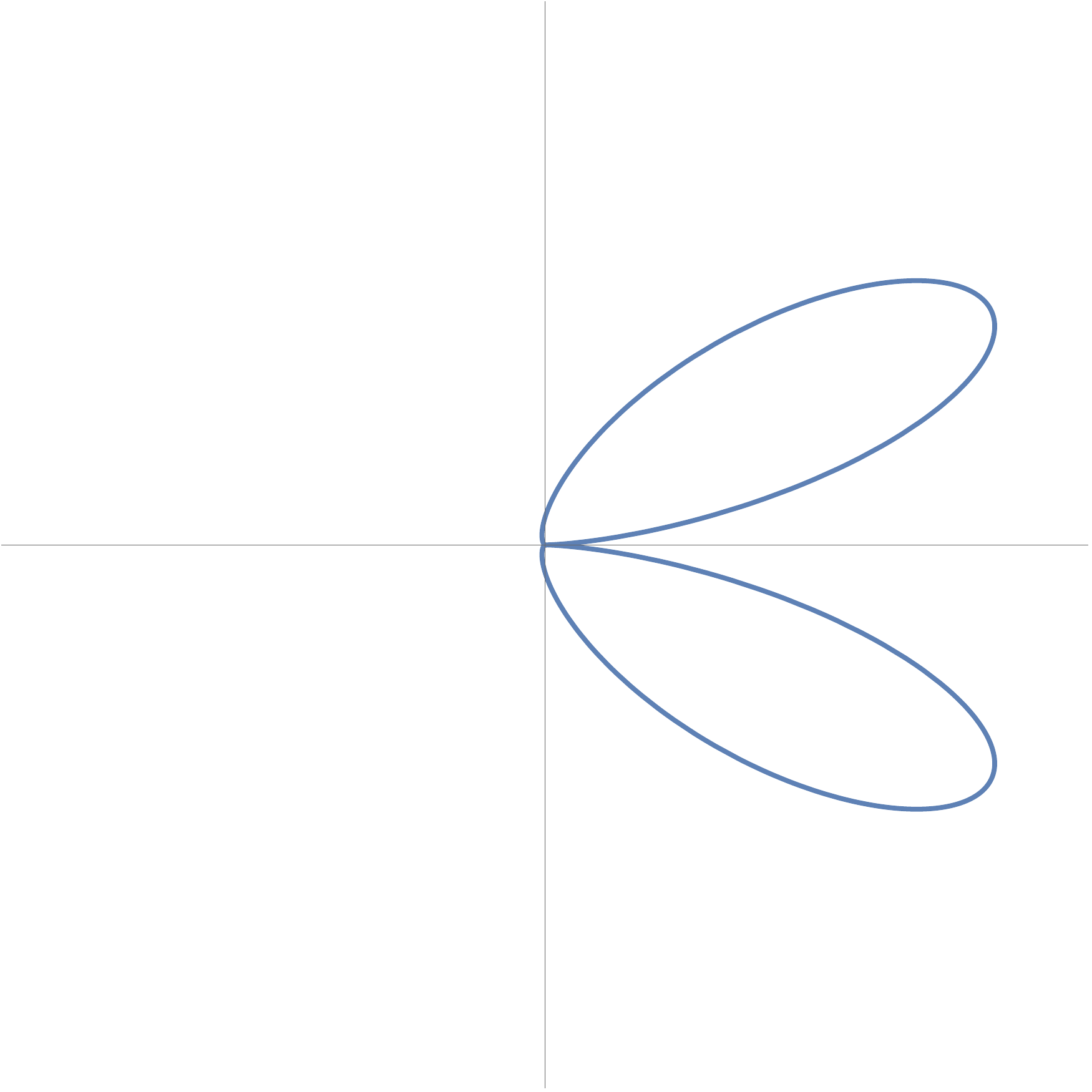}}\quad
\subfigure{\includegraphics[width=1.4in]{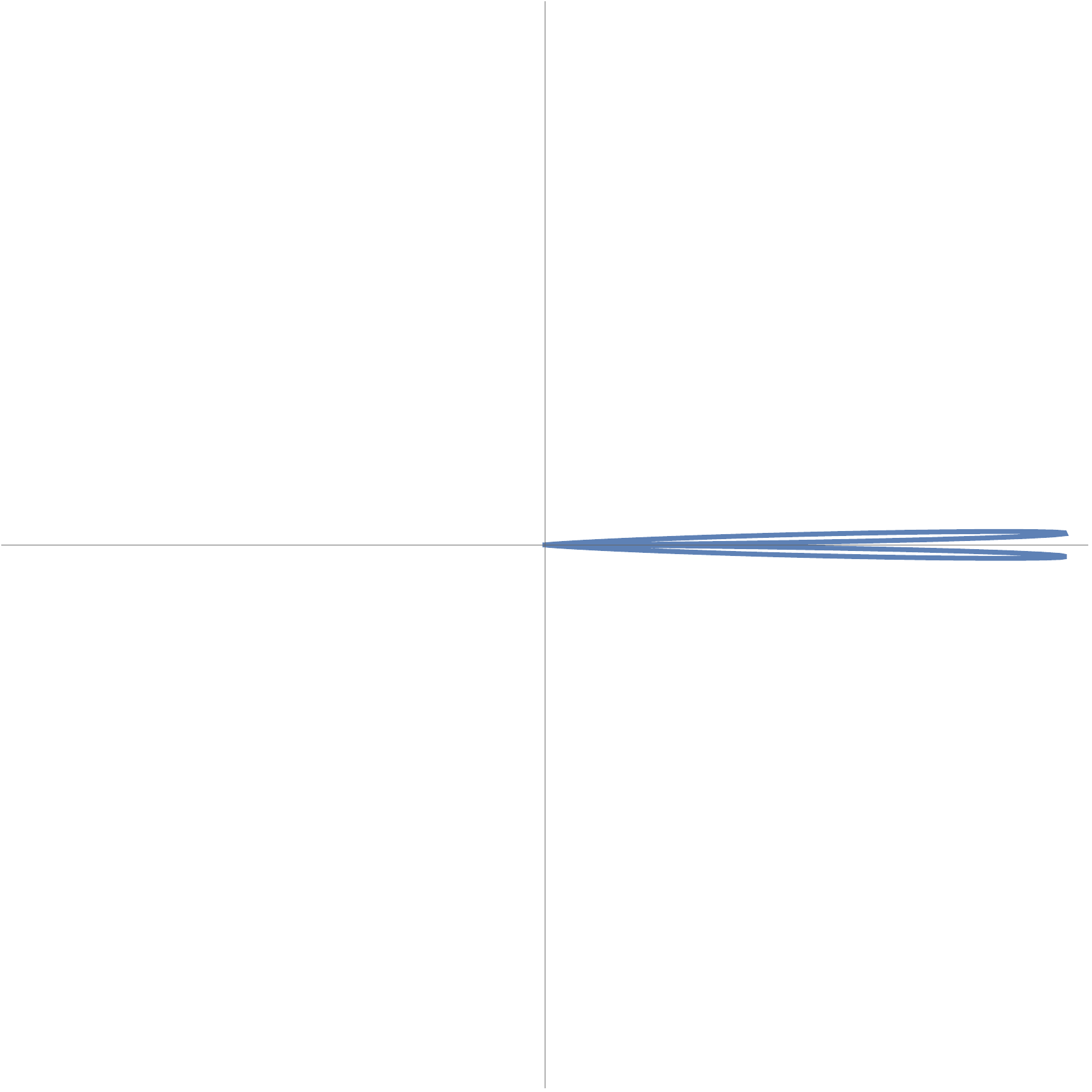} }}
\caption{\label{fig3} Rectilinear Angular Distribution; A polar plot with $\beta = 0.000, 0.333, 0.666, 0.999$. The vertical axes is $x$ and the horizontal axis is $z$. The electron moves forward along the $z$ straight line direction.   } 
\end{center}
\end{figure}

\begin{figure}[ht]
\begin{center}
\mbox{
\subfigure{\includegraphics[width=1.4in]{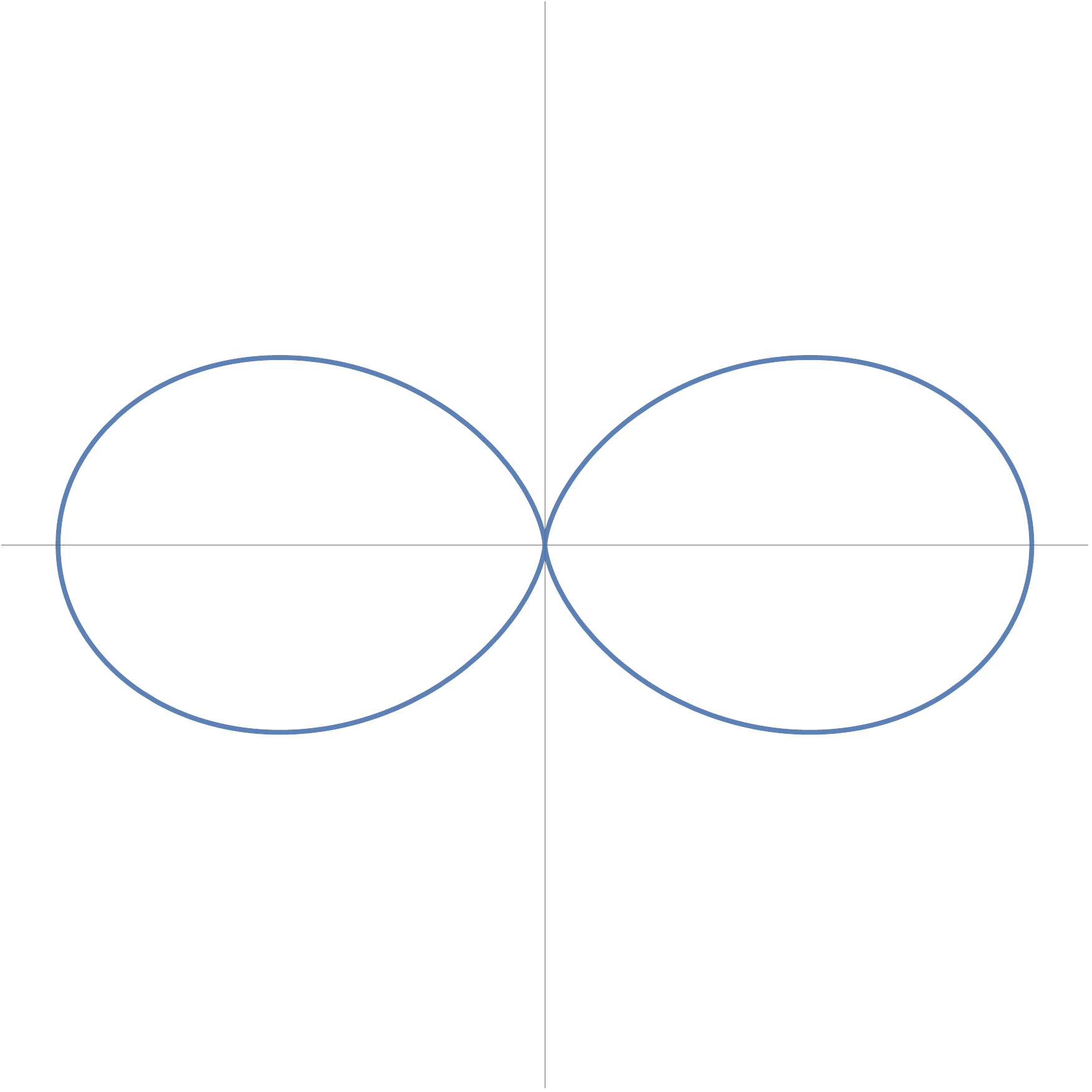}}\quad
\subfigure{\includegraphics[width=1.4in]{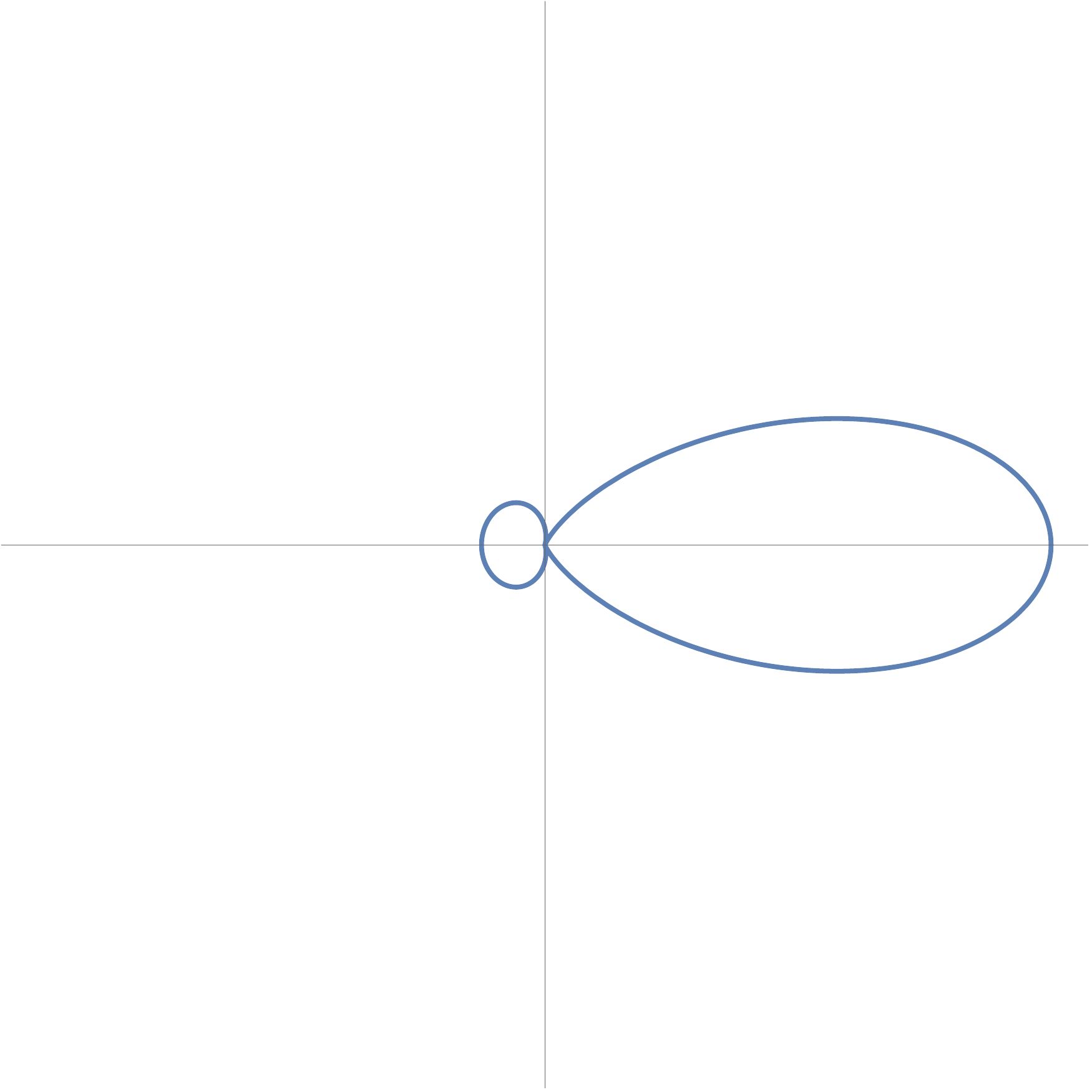}}\quad
\subfigure{\includegraphics[width=1.4in]{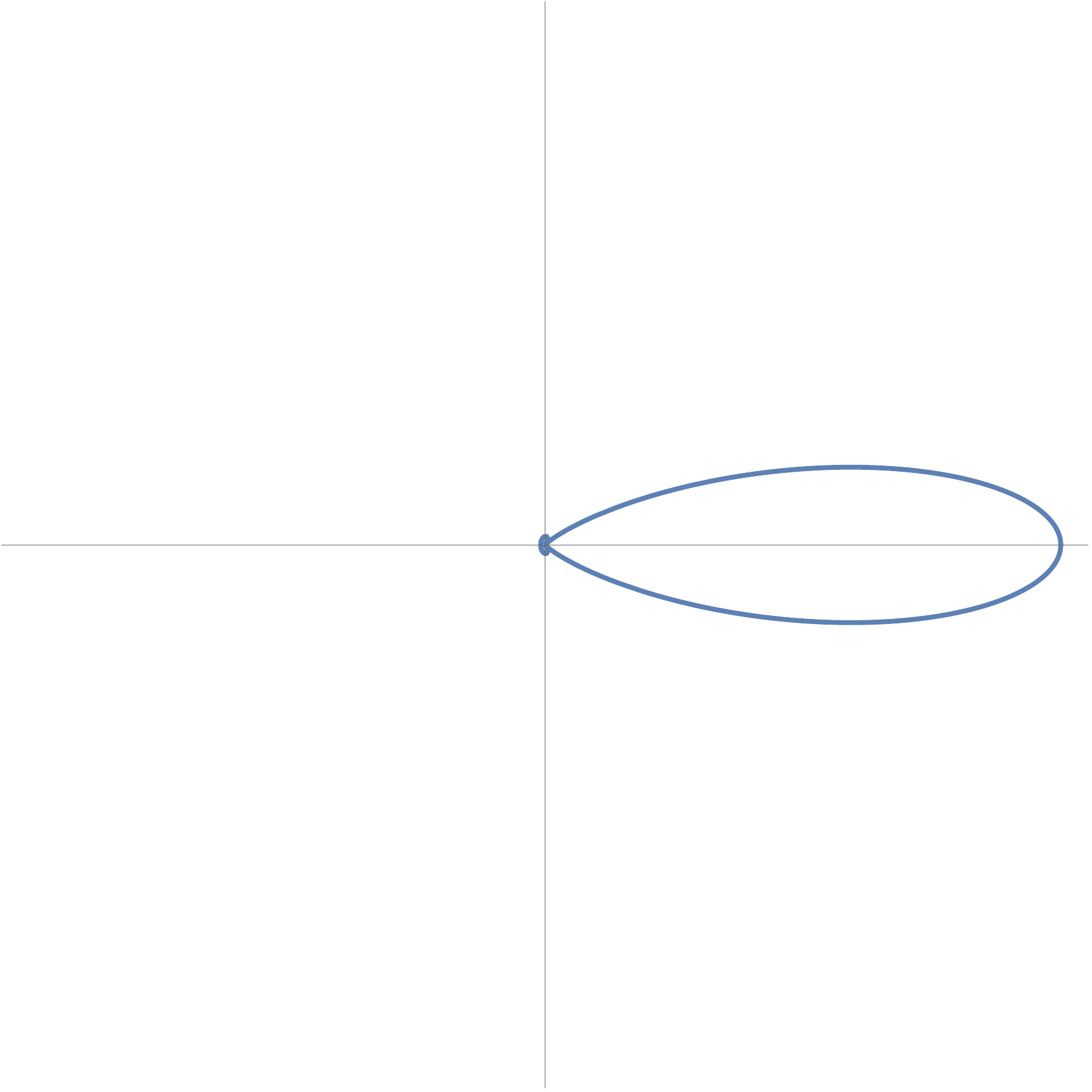}}\quad
\subfigure{\includegraphics[width=1.4in]{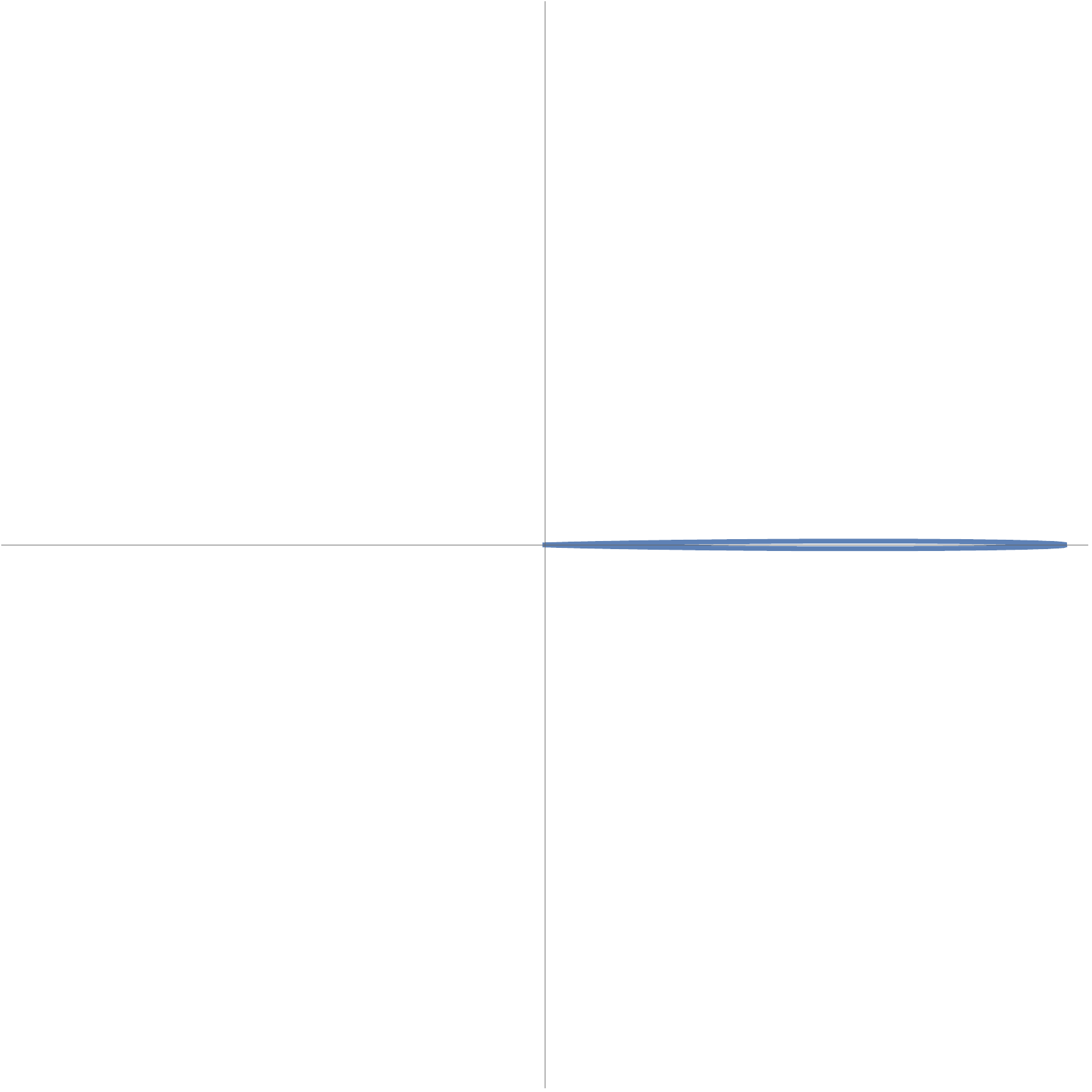} }}
\caption{\label{fig4} Synchrotron Angular Distribution; A polar plot with $\beta = 0.000, 0.333, 0.666, 0.999$. The vertical axes is $x$ and the horizontal axis is $z$. The electron moves forward in a circle toward the horizontal $z$ axis.} 
\end{center}
\end{figure}

\begin{figure}[ht]
\begin{center}
\mbox{
\subfigure{\includegraphics[width=1.4in]{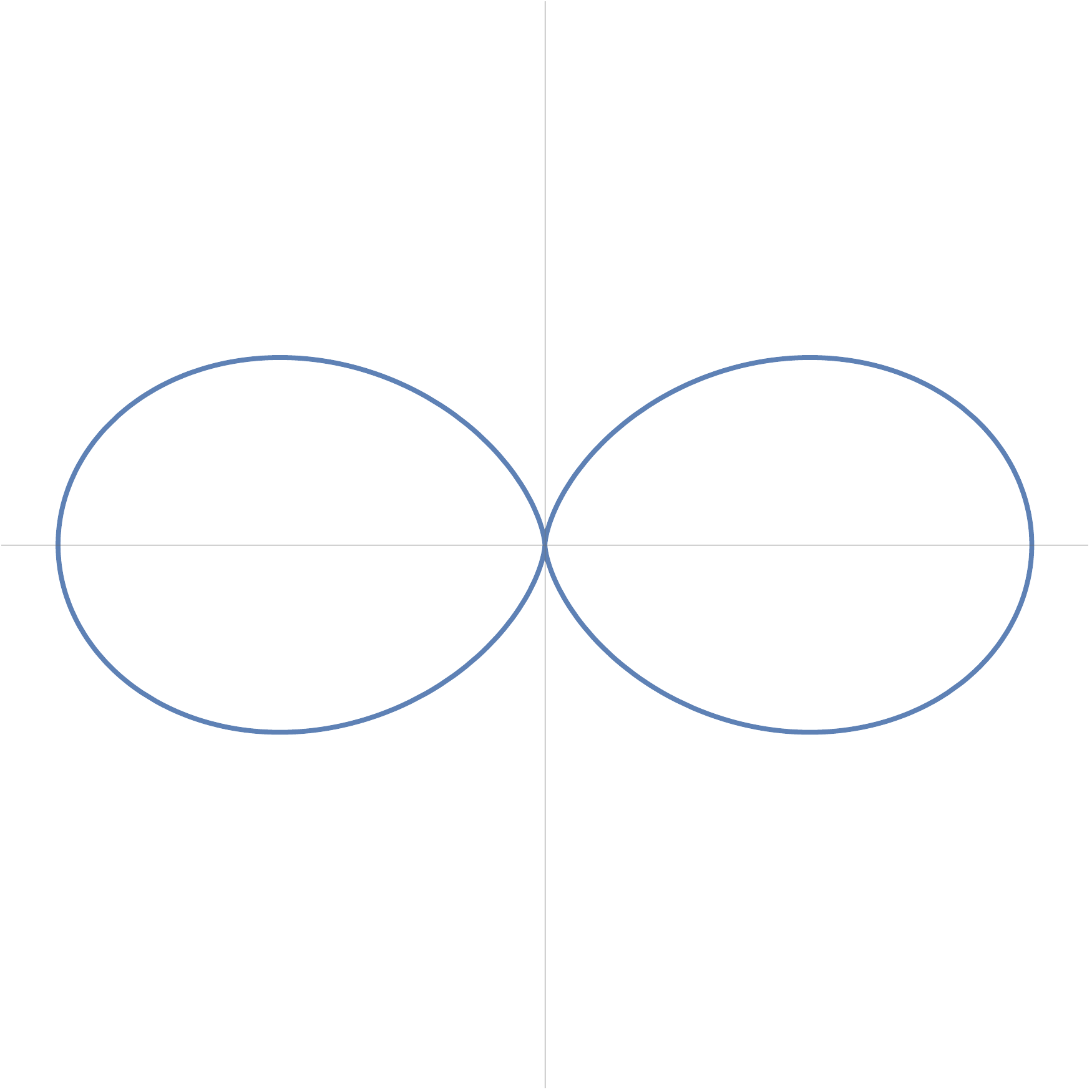}}\quad
\subfigure{\includegraphics[width=1.4in]{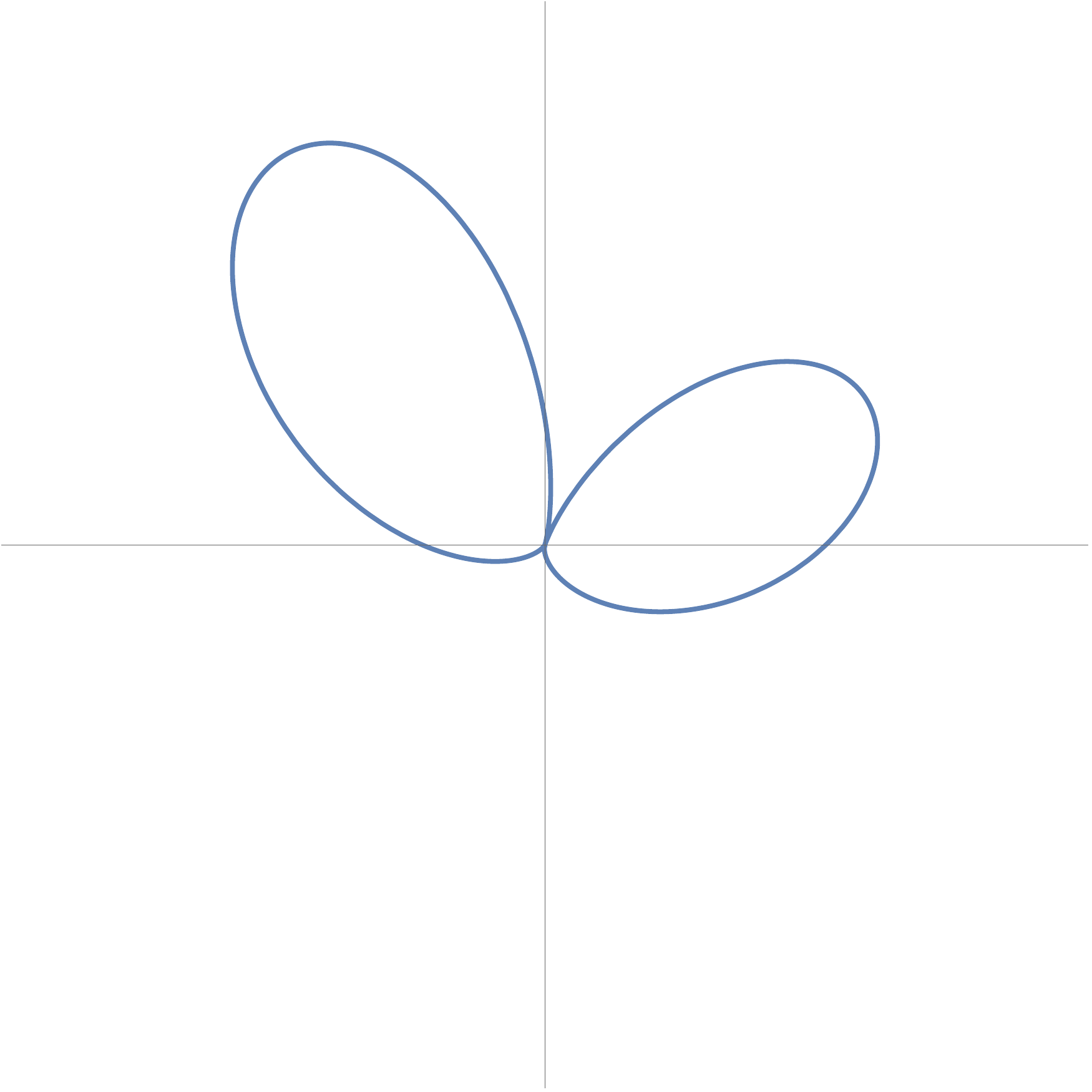}}\quad
\subfigure{\includegraphics[width=1.4in]{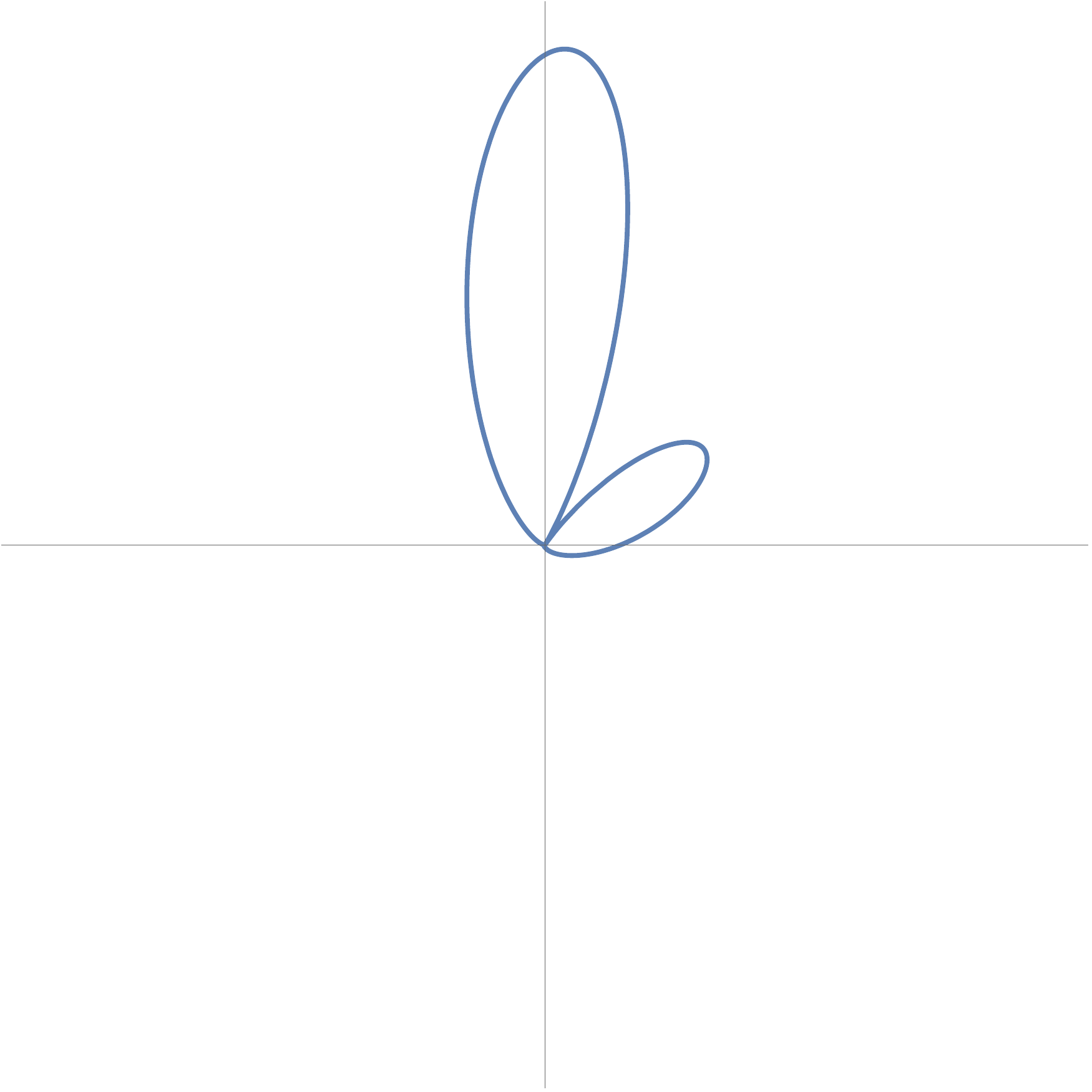}}\quad
\subfigure{\includegraphics[width=1.4in]{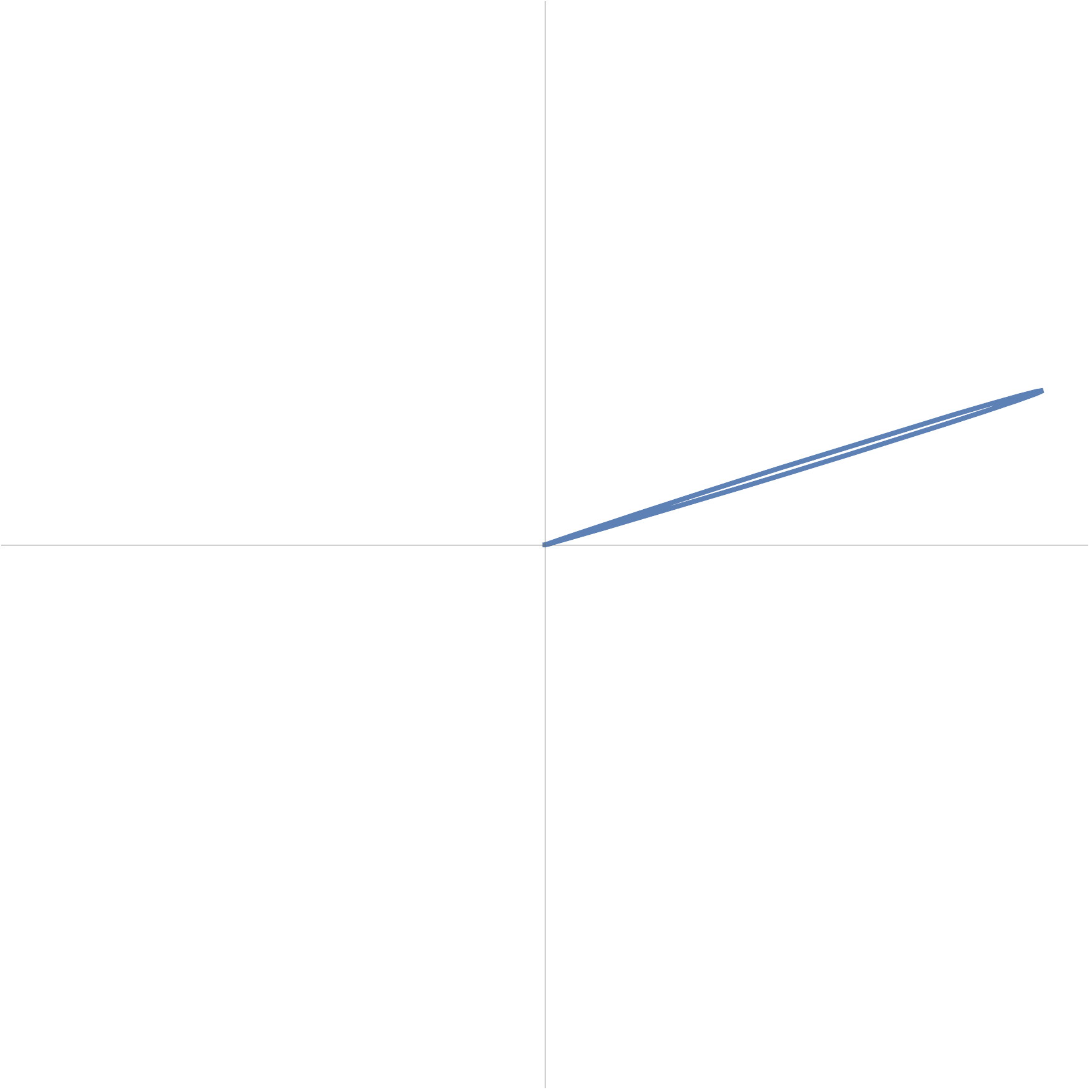} }}
\caption{\label{fig5} Cusp Angular Distribution; A polar plot with $\beta = 0.000, 0.333, 0.666, 0.999$. The vertical axes is $x$ and the horizontal axis is $z$. The electron moves in both dimensions.  } 
\end{center}
\end{figure}

At high speeds on the other hand, the situation is more interesting:

\begin{eqnarray}
\la \sin^2\theta\ra_{\textrm{line}} = \frac{1}{\gamma^2} + \mathcal{O}(\gamma^{-3})\,,\quad
\la \sin^2\theta\ra_{\textrm{circ}} = \frac{1}{\gamma^2} + \mathcal{O}(\gamma^{-3})\,,\quad
\la \sin^2\theta\ra_{\textrm{cusp}} = \frac{2}{\gamma} + \mathcal{O}(\gamma^{-2})\,.
\end{eqnarray}
%
The different scaling at high speeds quantifies the forward beaming angle as a function of speed.  This approach to $\theta_{\textrm{max}} =0$ demonstrates the need for much higher speeds for cusp motion to achieve forward beaming.  This can also be seen graphically for instance when at $99.9\%$ the speed of light the distribution of cusp radiation is significantly different from circular and rectilinear cases.  The distribution for the cusp motion is offset horizontally relative rectilinear, Fig.~(\ref{fig3}), and circular, Fig.~(\ref{fig4}), distributions, as can be seen in Fig.~(\ref{fig5}). These results agree with the speed requirements for $\theta^{\textrm{cusp}}_{\textrm{max}} \sim \gamma^{-1/2}$ scaling, Eq.~(\ref{maxangle}), relative to $\theta^{\textrm{line}}_{\textrm{max}} \sim \gamma^{-1}$ and $\theta^{\textrm{circ}}_{\textrm{max}} = 0$. 
%


\section{Conclusion}\label{sec:conclusions}

We have compared rectilinear radiation and synchrotron radiation to cusp radiation with a focus on both low and high speeds. We have found the acceleration ratio and the angular distribution of cusp radiation.  The cusp acceleration ratio is not as simple as the other uniformly accelerated motions and scales as a fraction of powers. Even though at low speeds we have determined cusp radiative power is more akin to rectilinear radiation, at high speeds the radiative power more closely scales like synchrotron radiation.  Interestingly, the angular distribution of the cusp power is more akin to synchrotron emission in both low and high speeds, at least to first order.
We analyzed the angular distribution for $\gamma \gg 1$ in part by calculating the maximum intensity in the forward direction, revealing the maximum angle scales differently for cusp radiation.  Identifying the intensity with synchrotron motion requires relaxing the restriction that circular motion be non-uniformly accelerated.  We have demonstrated the exact scaling requirements to achieve forward beaming for cusp radiation.  

\section*{Acknowledgment}
MRRG thanks the Julian Schwinger Foundation for its support under Grant 15-07-0000.

\newpage
\section*{Appendix A: Details to Determine the Angular Distribution}

The necessary tools for determining the angular distribution, Eq.~(\ref{angular1}),
\be\label{AngDistr}  \frac{dP}{d\Omega} = \frac{q^2}{4\pi} \frac{|\hrcurs \times (\bd{u} \times \bd{a})|^2}{(\hrcurs \cdot \bd{u})^5}, \ee
is shown in the following.  Here 
\be \hrcurs \times(\bd{u} \times \bd{a}) = (\hrcurs \cdot \bd{a})\bd{u} - (\hrcurs \cdot \bd{u})\bd{a}, \ee
so that
\be  |\hrcurs \times (\bd{u} \times \bd{a})|^2 = (\hrcurs \cdot \bd{a})^2 u^2 - 2 (\bd{u}\cdot \bd{a})(\hrcurs \cdot \bd{a})(\hrcurs \cdot \bd{u}) + (\hrcurs \cdot \bd{u})^2 a^2. \ee
Using the definitions, including $\brcurs \equiv \bd{r} - \bd{r}'$, (the vector from a source point $\bd{r}'$ to a field point $\bd{r}$), where the source point is our origin, $\bd{r}'=0$,
\be \hrcurs \equiv \frac{\bd{r} - \bd{r}'}{|\bd{r} - \bd{r}'|} = \bd{\hat{r}} \equiv \sin\theta\cos\phi \bd{\hat{x}} + \sin\theta\sin\phi \bd{\hat{y}} + \cos\theta \bd{\hat{z}},\ee
\be \bd{u} \equiv \hrcurs - \bd{v}, \ee
one finds the terms in the order that they appear,
\be \hrcurs \cdot \bd{a} = a_x \sin\theta\cos\phi + a_y\sin\theta\sin\phi + a_z \cos\theta, \ee
\be u^2 = (\hrcurs - \bd{v})^2 = 1 - 2( v_x \sin\theta\cos\phi + v_y\sin\theta\sin\phi + v_z \cos\theta) + v^2,\ee
\be \bd{u}\cdot \bd{a} = (\hrcurs - \bd{v})\cdot \bd{a} = a_x \sin\theta \cos\phi + a_y\sin\theta\sin\phi + a_z \cos\theta -v_x a_x -v_y a_y - v_z a_z, \ee
\be \hrcurs \cdot \bd{u} = \hrcurs^2 - \hrcurs\cdot \bd{v} = 1- v_x \sin\theta\cos\phi -v_y\sin\theta\sin\phi - v_z \cos\theta, \ee
\be a^2 = a_x^2 + a_y^2 + a_z^2, \quad v^2 = v_x^2 + v_y^2 + v_z^2. \ee
The cusp values for the component velocities and accelerations are,
\be a_x = a\frac{2-\gamma}{\sqrt{\gamma^2 - 2 \gamma + 2}}, \quad a_y = 0, \quad a_z = a\frac{\sqrt{2(\gamma-1)}}{\sqrt{\gamma^2-2\gamma+2}},\ee
\be v_x = \gamma^{-1}\sqrt{2(\gamma-1)}, \quad v_y = 0, \quad v_z = 1-\gamma^{-1}.\ee
\newpage

\newpage
\subsubsection*{Plots of Angular Distribution}

\newpage

\end{document}